\documentclass[preprint,superscriptaddress,amsmath,amssymb]{revtex4}
\usepackage{graphicx,color,slashed,bbold}

\usepackage{slashed,bbold}
\usepackage{soul}

\begin{document}

{\begin{flushright}{KIAS-P17025}
\end{flushright}}

\title{ $L_\mu -L_\tau$ gauge-boson production from lepton flavor violating $\tau$ decays at Belle II }

\author{Chuan-Hung Chen}
\email{physchen@mail.ncku.edu.tw}
\affiliation{Department of Physics, National Cheng-Kung University, Tainan 70101, Taiwan}

\author{Takaaki Nomura}
\email{nomura@kias.re.kr}
\affiliation{School of Physics, KIAS, Seoul 02455, Korea}

\date{\today}

\begin{abstract}

$L_\mu -L_\tau$ gauge boson ($Z'$) with a mass in the MeV to GeV region  can resolve not only the muon $g-2$ excess, but also the gap in the high-energy cosmic neutrino spectrum at IceCube. It was recently proposed that such a light gauge boson can be detected during the Belle II experiment with a luminosity of 50 ab$^{-1}$ by the $e^+ e^- \to \gamma +\slashed{E}$ process through the kinetic mixing with the photon, where the missing energy $\slashed{E}$ is from the $Z' \to \bar\nu \nu$ decays. We study the phenomenological implications when a pair of singlet vector-like leptons carrying different $L_\mu - L_\tau$ charges  are included,  and a complex singlet scalar ($\phi_S$) is introduced to accomplish the spontaneous $U(1)_{L_\mu -L_\tau}$ symmetry breaking. It is found that the extension leads to several phenomena of interest, including (i) branching ratio (BR) for $h\to \mu \tau$ can be of the order of $10^{-3}$; (ii) $\phi_S$-mediated muon $g-2$ can be of the order of $10\times 10^{-10}$; (iii) BR for $\tau \to \mu \phi^*_S\to \mu  Z'Z'$ can be $10^{-8}$, and (iv) kinetic mixing between the $Z'$ boson and photon is sensitive to the relative heavy lepton masses.  The predicted BRs for $\tau\to (3\mu+\slashed{E}, 5\mu$)  through the leptonic $Z'$  decays can reach a level of $10^{-9}$, in which the results fall within the sensitivity of the Belle II in the search for  the rare tau decays. 

\end{abstract}

\maketitle

\section{Introduction}

A $Z'$ gauge boson, dictated by an anomaly-free $U(1)_{L_\mu -L_\tau}$ gauge symmetry~\cite{He:1990pn,Foot:1994vd}, has been broadly studied. Especially, the $Z'$ boson with a mass in the range between  MeV and GeV can help explain  observed anomalies, such as the muon anomalous magnetic moment (muon $g-2$)~\cite{Altmannshofer:2016brv,Gninenko:2014pea,Gninenko:2001hx}, and deficiencies in the high-energy cosmic neutrino spectrum reported by IceCube~\cite{Aartsen:2014gkd,Araki:2014ona,Kamada:2015era,DiFranzo:2015qea,Araki:2015mya}. In addition, the $U(1)_{\mu-\tau}\equiv U(1)_{L_\mu -L_\tau}$ gauge model can be also used to resolve the largely unexpected  lepton-flavor nonuniversality in semileptonic $B$ decays~\cite{Altmannshofer:2014cfa,Crivellin:2015mga,Altmannshofer:2016jzy}, Higgs $h$ lepton flavor violating (FLV) decays~\cite{Crivellin:2015mga,Heeck:2014qea,Altmannshofer:2016oaq,Heeck:2016xkh}, and  dark matter and/or neutrino mass~\cite{Baek:2015fea,Biswas:2016yan,Patra:2016shz,Baek:2015mna,Lee:2017ekw}. 

Recently, there has been some progress made with detecting the light $Z'$ boson in experiments and  limiting the  ranges of $Z'$ mass  $m_{Z'}$ and  gauge coupling $g_{Z'}$, which are used to fit the muon $g-2$ anomaly.   For instance, according to neutrino trident production processes, which were measured by the CHARM-II collaboration~\cite{Geiregat:1990gz} and CCFR collaboration~\cite{Mishra:1991bv},  it was shown that  $m_{Z'}\gtrsim 400$ MeV and $g_{Z'} > {\rm few}\times  10^{-3}$ are excluded~\cite{Altmannshofer:2014pba}.  Based on a observation  of  cross-section of the $e^+ e^- \to \mu^+ \mu^- Z'$, $Z'\to \mu^+ \mu^-$ channel, which was recently  measured  by the BABAR collaboration at a 90\% confidence level (CL)~\cite{TheBABAR:2016rlg}, the bound of $g_{Z'} < 0.7\times 10^{-3}$ at $m_{Z'}\approx 0.22$ GeV can be obtained. The  ranges of  $m_{Z'} \subset (1,10)$ MeV and $g_{Z'}\subset (0.1,1)\times 10^{-3}$ were badly  narrowed~\cite{Harnik:2012ni} by the measurement of $^7Be$ solar neutrino scattering off the electron  in the Borexino experiment~\cite{Bellini:2011rx}, where the $\nu$-$e$ scattering occurred through loop-induced kinetic mixing between the electromagnetic and $Z'$ gauge fields.  Although the $m_{Z'}$ and $g_{Z'}$ parameter spaces for explaining the muon $g-2$ excess are not completely excluded, the allowed ranges are  strictly bounded by the experiments above. 

A detection of the light $Z'$ gauge boson via the process $e^+ e^- \to \gamma + \slashed{E}$ at the Belle II, which will record an unprecedented data sample of 50 ab$^{-1}$, was recently proposed in~\cite{Araki:2017wyg,Kaneta:2016uyt}, where  $\slashed{E}$ is the missing energy from the $Z' \to \bar\nu \nu$ decays, and the $Z'$ boson is produced through  kinetic mixing  with the photon. Although kinetic mixing was also involved in the Borexino  $\nu$-$e$ scattering experiment, it was found that the loop-induced mixing in the $e^+ e^- \to \gamma Z'$ process  is dependent on the $q^2=m^2_{Z'}$, whereas the mixing in the solar neutrino experiment is a constant in  $q^2$ due to the low energy neutrinos.  The kinetic mixing parameters in the both processes are respectively written as~\cite{Araki:2017wyg}:
 \begin{align}
 \epsilon_{\rm Belle} & = \frac{ e g_{Z'}}{2\pi^2} \int^1_0 dx\, x(1-x) \ln\frac{m^2_\tau - x(1-x) q^2 }{m^2_\mu - x(1-x) q^2}\,, \quad \epsilon_{\nu e} = \frac{e g_{Z'}}{6\pi^2} \ln\frac{m_\tau}{m_\mu}\,. \label{eq:ep_Belle}
 \end{align}
It has been concluded that with a Belle-II integrated luminosity of 50 ab$^{-1}$, the significance of  an $e^+ e^- \to \gamma +\slashed{E}$ process higher  than $3\sigma$ significance  can be reached, where the sensitive regions are  $m_{Z'} \lesssim 1$ GeV and $g_{Z'} \gtrsim 0.8 \times 10^{-3}$. 

If we  examine the light $Z'$ gauge boson together with the spontaneous $U(1)_{\mu-\tau}$ symmetry breaking method, it can be found that in addition to the $m_{Z'}$ and $g_{Z'}$ parameters, the $U(1)_{\mu-\tau}$ gauge model 
needs at least one more new free parameter to dictate the mass of a scalar boson, in which the scalar field only carries the $U(1)_{\mu-\tau}$ charge and is  responsible for  the  symmetry breaking. If we employ a complex singlet scalar field ($\phi_S$) to accomplish the symmetry breaking, the $\phi_S$-$Z'$-$Z'$ coupling from the kinetic term can lead to  the $\phi_S \to Z' Z'$ decay when $m_S > 2 m_{Z'}$ is satisfied. If the singlet scalar can be produced with a sizable cross section, the light $Z'$ can then be generated through the $\phi_S$ decay. However, if the $\phi_S$ field only couples to the leptons  via the Yukawa interactions,   we cannot have the $SU(2)_L \times U(1)_Y \times U(1)_{\mu-\tau}$ gauge invariant Yukawa couplings because the left-handed and right-handed leptons are  $SU(2)_L$ doublets and singlets, respectively.  Thus, it is difficult to generate the singlet scalar boson and detect the $Z'$ signal through the $\phi_S\to Z' Z'$ channel.  

We find that if two singlet vector-like  leptons, which  carry different $U(1)_{\mu-\tau}$ charges, are added to the model,    probing the light $Z'$ through $\phi_S \to Z' Z'$  can then be achieved.  The resulting model not only  is   $U(1)_Y$ and $U(1)_{\mu-\tau}$ gauge anomaly-free, but it also removes the scale dependence of  loop-induced kinetic mixing.  As a consequence, several phenomena of interest are induced, including (i)  LFV branching ratios (BRs) for  the $h, \phi_S \to \mu \tau$ decays can be of the order of $10^{-3}$;  (ii) muon $g-2$  from the same LFV effects can achieve a level of $10^{-9}$; (iii) BR for $\tau\to \mu \phi^*_S \to \mu Z' Z'$ can be of the  order of $10^{-8}$;  (iv) the kinetic mixing of Eq.~(\ref{eq:ep_Belle}) is modified and becomes sensitive to the relatively heavy lepton masses.

With 50 ab$^{-1}$ of data accumulated at the Belle II, the sample of $\tau$ pairs can be increased up to around $5\times 10^{10}$, where the sensitivity necessary to observe the LFV $\tau$ decays can reach $10^{-10}-10^{-9}$, depending on the processes~\cite{Aushev:2010bq}.  If $m_{Z'}> 2 m_{\mu}$ and  $BR(Z' \to \mu^+ \mu^-)\times BR(Z' \to \bar\nu \nu, \mu^+ \mu^+) \sim 0.2$,  we will show that the BRs for the $\tau \to 3 \mu + \slashed{E}$ and $\tau \to 5\mu$ decays  can be ${\cal O}(10^{-9})$ in the extension of the SM. Intriguingly, the resulting BRs of the new tau decay channels are located in the Belle II sensitivity. Since $\tau \to 3\mu$ and $\tau \to (e ,\mu) \gamma$ are suppressed in this model, the detectable $\tau \to 3 \mu + \slashed{E}$ and $\tau \to 5\mu$ decays can be used as the characteristics that distinguish  them  from other models, which have sizable BRs for the $\tau \to 3\mu$ and $\tau\to (e, \mu) \gamma$ decays. 

The E821 experiment at the Brookhaven National Laboratory (BNL)~\cite{Bennett:2006fi} revealed the  uncertainty of the measured muon $g-2$ to be 0.54 ppm, and a result of over a $3\sigma$ deviation from the SM prediction was obtained. The new muon $g-2$ measurements performed in the  E989 experiment at Fermilab and the E34 experiment at J-PARC will aim for   a precision  of 0.14 ppm~\cite{Grange:2015fou} and 0.10 ppm~\cite{Otani:2015jra}, respectively.  Thus,   the muon $g-2$ induced by the LFV effects  in this model can be strictly bounded with more accurate  measurements. Hence,  in this work, we plan to show the impacts on lepton-flavor conservation and the LFV phenomena  when  $U(1)_{\mu-\tau}$ gauge symmetry and two singlet vector-like leptons are introduced to the SM.

The paper is organized as follows. In Sec.~II, we introduce the $U(1)_{\mu-\tau}$ extension of the SM by adding two singlet vector-like leptons. The new Yukawa, $Z$, and $Z'$ couplings are derived in this section. We show the numerical analysis on the phenomena of interest  in Sec.~III, where they include $h\to \mu \tau$, muon $g-2$, rare $\tau$ decays, and the influence on the $e^+ e^-\to \gamma Z'$. The summary is given in Sec.~IV. 

\section{Gauged $L_\mu -L_\tau$ Model}

In the following, we begin to introduce the new interactions in the extension of the SM. In a gauged $L_\mu - L_\tau$ model,  we add two singlet vector-like leptons ($\ell_{4,5}$) and a complex singlet  scalar field ($S$) into the SM, where the $S$ field is responsible for the  spontaneous   $U(1)_{\mu-\tau}$  symmetry breaking, and the heavy leptons lead to lepton-flavor changing neutral currents (LFCNCs) through the Yukawa couplings. In order to obtain the Higgs lepton-flavor violation and remove the scale dependence of the loop-induced kinetic mixing between the photon  and the $Z'$ gauge boson, the $U(1)_{\mu-\tau}$ charges of $\ell_4$ and $\ell_5$ must be opposite in sign. When the charge of $\ell_4$  is determined, the  charge of $S$ then is certain.  For clarity, we show the $U(1)_{\mu-\tau}$ charges of the leptons and $S$ field in Table~\ref{tab:U1}. Accordingly,   the  Yukawa interactions, which satisfy the $SU(2)_L\times U(1)_Y \times U(1)_{\mu -\tau}$ gauge symmetry, are written as: 
 \begin{align}
 -{\cal L}_Y & = Y_{\ell} \bar L_{\ell} H \ell_R + y_\mu \bar L_\mu H \ell_{4 R} + y_\tau \bar\ell_{4L} \tau_R S + m_{4L} \bar \ell_{4L} \ell_{4R}  + y'_\tau \bar L_\tau H \ell_{5 R}   \nonumber \\
& + y'_\mu \bar\ell_{5L} \mu_R S^\dagger + y_S \bar \ell_{4 L} \ell_{5 R} S + y'_S \bar \ell_{5L} \ell_{4R} S^\dagger + m_{5L}  \bar \ell_{5L} \ell_{5R} + H.c. \,,  \label{eq:Yukawa}
  \end{align} 
where $L_\ell$ denotes the SM doublet lepton, $f_{R(L)} = P_{R(L)} f $ with $P_{R(L)}=(1\pm \gamma_5)/2$; $H$ is the SM Higgs doublet, and $m_{4L,5L}$ are the heavy lepton masses. The electroweak and $U(1)_{\mu-\tau}$ symmetries can be spontaneously broken through $\langle H \rangle = (v +h)/\sqrt{2}$ and $\langle S \rangle = (v_S + \phi_S) /\sqrt{2}$, where $v (v_S)$ is the vacuum expectation value (VEV) of the $H(S)$ field.  From Eq.~(\ref{eq:Yukawa}), it can be seen that $\ell_4$ and $\ell_5$ can mix together through the $y_S \bar\ell_4 \ell_5 S$ term when the $U(1)_{\mu-\tau}$ symmetry is  broken. In order to simplify the following formulation, we assume $y'_S=y_S$ and take the basis, $\ell'_4 = \cos\alpha \ell_ 4 - \sin\alpha \ell_5$ and $\ell'_5 = \sin\alpha \ell_4 + \cos\alpha \ell_5$, so that  the $2\times 2$ mass matrix of $\ell_4$ and $\ell_5$ is diagonalized as:
 \begin{equation}
 \left( \begin{array}{cc}
   m_{L1} & 0  \\
 0   &   m_{L2}
 \end{array} \right)
   =   \left( \begin{array}{cc}
   \cos\alpha & -\sin\alpha  \\
  \sin\alpha   &   \cos\alpha
   \end{array} \right)
    \left( \begin{array}{cc}
   m_{4L} & \frac{y_S v_S}{\sqrt{2}}  \\
 \frac{y_S v_S}{\sqrt{2}}   &   m_{5L}
 \end{array} \right)
 \left( \begin{array}{cc}
   \cos\alpha & \sin\alpha  \\
 - \sin\alpha   &   \cos\alpha
   \end{array} \right)\,,
 \end{equation}
where  the  $m_{L1}$, $m_{L2}$, and the mixing angle $\alpha$ can be related  to  the $m_{4L,5L}$ and $y_S v_S$ parameters  as:
 \begin{align}
m_{L1,L2} & = \frac{1}{2} \left( m_{4L} + m_{5L} \pm \sqrt{(m_{5L} - m_{4L})^2 + 2 y^2_S v^2_S}\right) \,, \nonumber \\
\tan2\alpha & = \frac{\sqrt{2} y_S v_S }{m_{5L} - m_{4L}}\,.
\end{align} 
We note that  in general, the SM Higgs can mix with the scalar  $\phi_S$ via the scalar potential. Since the mixing  is a new free parameter, in order to avoid the constraint resulting from the precision Higgs measurements, hereafter, we consider the mixing to be small and neglect its contributions.  
\begin{table}[tb]
\caption{$U(1)_{\mu-\tau}$ charges of leptons and $S$ field.}
\begin{tabular}{ccccccc} \hline \hline
            &  ~~~$e$~~~  & ~~~$\mu$~~~ & ~~~ $\tau$ ~~~& ~~~ $\ell_4$  ~~~ & ~~~ $\ell_5$  ~~~& ~~~$S$~~~\\ \hline 
 $U(1)$ &   0   &   1      & $-1$      &  1  & $-1$ & $2$\\ \hline \hline

\end{tabular}
\label{tab:U1}
\end{table}%

 Since the model involves a new scalar field $S$, to understand its properties, we write the gauge invariant  scalar potential   as:
 \begin{align}
 V =&  - \mu_H^2 H^\dagger H - \mu_S^2 S^\dagger S + \lambda_H (H^\dagger H)^2 + \lambda_S (S^\dagger S)^2 + \lambda_{HS} (H^\dagger H)(S^\dagger S)\,
 \end{align}
 where $\mu^2_{H,S}$ and $\lambda_{H,S}$ are positive parameters.  Based on the minimum condition, the VEVs of the scalar fields $H$ and $S$ can be obtained as:
 \begin{align}
 \frac{\partial V}{\partial v} &= v \left( - \mu_H^2 + \lambda_H v^2 + \frac{1}{2} \lambda_{HS} v_S^2 \right) =0,  \nonumber \\
  \frac{\partial V}{\partial v_S} &= v_S \left( - \mu_S^2 + \lambda_S v_S^2 + \frac{1}{2} \lambda_{HS} v^2 \right) =0.
 \end{align}
With the assumption of $\lambda_{HS} \ll 1$, we obtain $v \simeq \sqrt{\mu_H^2/\lambda_H}$ and $v_S \simeq \sqrt{\mu_S^2/\lambda_S}$.
 From the scalar potential, the mass-squared  matrix for the $h$ and $\phi_S$ is expressed as:  
 \begin{equation}
 M_{\phi}^2 = \begin{pmatrix} 2 \mu_H^2 & \lambda_{HS} v v_S \\ \lambda_{HS} v v_S & 2 \mu_S^2 \end{pmatrix}. \label{eq:M2phi}
 \end{equation}
 The  eigenvalues and eigenstates of Eq.~(\ref{eq:M2phi}) are then obtained as:
 \begin{align}
&  m^2_{H_1,H_2}=  (\mu_H^2 + \mu_S^2) \pm  \sqrt{(\mu_H ^2- \mu_S^2)^2 +  \lambda_{HS} v v_S}, \nonumber \\
& \begin{pmatrix} H_1 \\ H_2 \end{pmatrix} = \begin{pmatrix} \cos \phi & \sin \phi \\ - \sin \phi & \cos \phi \end{pmatrix} \begin{pmatrix} \phi_S \\ h \end{pmatrix}, \quad
\tan 2 \phi = \frac{\lambda_{HS} v v_S }{\mu_H^2 - \mu_S^2},
 \end{align}
 where $H_2$ is the SM-like Higgs boson; and the mass hierarchy $m_{H_1} < m_{H_2}$ is assumed in this paper.  The mixing angle $\phi$ can be  constrained by the  SM Higgs precision measurements. Especially, when $m_{H_2}> 2 m_{H_1} , 2m_{Z'}$, the $H_2 \to H_1 H_1$ and $H_2 \to Z' Z'$ decays will be opened.  
 Since the mixing angle $\phi$ is irrelevant to our study, in the following analysis, we take $\phi\ll 1$,  $H_2 \simeq h$, and $H_1\simeq \phi_S$.  As a result, we have $v_S \simeq  m_{S}/\sqrt{2\lambda_S }$; that is, $v_S \sim 100$ GeV and $m_S\sim 10$ GeV can be achieved when $\lambda_S \sim 5\times 10^{-3}$.

 In order to obtain the $Z'$ mass and the $S$ gauge coupling to the $Z'$ boson, we write the covariant derivative of the $S$ field to be $D_\mu = \partial_\mu + i g_{Z'} X_S Z'_\mu$, where $X_S=2$ is the $U(1)_{\mu-\tau}$ charge of the $S$ field.  With $\langle S \rangle = (v_S+ \phi_S)/\sqrt{2}$,   the $Z'$ mass and $\phi_S-Z'-Z'$ coupling can be obtained through the kinetic term  as:
 \begin{align}
 (D^\mu S)^\dagger  (D_\mu S) & \supset \frac{1}{2} ( v_S + \phi_S)^2  g^2_{Z'} X^2_S Z'_\mu Z'^{\mu} \,, \nonumber \\
  m_{Z'}&= 2 g_{Z'} v_S\,,\quad
  \phi_S -Z'-Z' : \frac{2 m^2_{Z'}}{v_S} g_{\mu \nu}\,. \label{eq:SZ'Z'}
 \end{align}
 
 After electroweak and $U(1)_{\mu-\tau}$ symmetry breaking, the lepton-flavor mixing information can be obtained from Eq.~(\ref{eq:Yukawa}). 
 If we combine the SM leptons and the heavy leptons  to form a multiplet state in flavor space, denoted by $ \ell'^T= (\pmb{\ell\,, \Psi}_\ell )$ with $\pmb{\ell}=(e, \mu,\tau)$ and $\pmb{\Psi}^T_{\ell'}=( \ell'_4, \ell'_5)$, the $5\times 5$ lepton mass matrix can be written as:
  \begin{equation}
  \bar\ell'_L  M_{\ell'}  \ell'_R =  \left(\begin{array}{cc}
 \pmb{ \bar \ell}_L \,, & \pmb{\bar \Psi}_{\ell L}  \end{array} \right) 
    \left( \begin{array}{c|c}
  ~~~{\pmb{m}_\ell}_{3\times 3}~~~ &  \pmb{ \delta m}_1  \\ \hline
 \pmb{\delta m}^T_2  &  \pmb{m}_L
 \end{array} \right)_{5\times 5} 
  \left(\begin{array}{c}
   \pmb{\ell}_R \\
   \pmb{\Psi}_{\ell R} \end{array}\right) \,, \label{eq:mass}
  \end{equation}
  where diag$\pmb{m_\ell}=( m_e, m_\mu, m_\tau)$, $m_f = v Y_f/\sqrt{2}$, diag$\pmb{m}_L = (m_{L1}, m_{L2})$,  and $\pmb{\delta m}_{1,2}$  are given by:
  \begin{align}
  \pmb{\delta m}^T_1 & = \left(\begin{array}{ccc}
 0 \,, & \frac{v y_\mu }{\sqrt{2}}c_\alpha\,, &  - \frac{v y'_\tau}{\sqrt{2}} s_\alpha \\
  0\,, &   \frac{v y_\mu }{\sqrt{2}}s_\alpha\,, & \frac{v y'_\tau}{\sqrt{2}} c_\alpha
 \end{array} \right)  \,, ~~
  \pmb{\delta m}^T_2 = \left(\begin{array}{ccc}
 0 \,, & -\frac{v_S y'_\mu }{\sqrt{2}}s_\alpha\,, &  \frac{v_S y_\tau }{\sqrt{2}} c_\alpha\\
  0\,, & \frac{v_S y'_\mu }{\sqrt{2}}c_\alpha \,, & \frac{v_S y_\tau }{\sqrt{2}} s_\alpha
 \end{array} \right)
     \end{align}
     with $c_\alpha=\cos\alpha$ and $s_\alpha=\sin\alpha$. 
    To diagonalize the mass matrix $M_{\ell'}$ in Eq.~(\ref{eq:mass}), we introduce two unitary matrices $V_{R, L}$. Since we take the flavor mixing effects to be perturbative and are suppressed by $m_{L1,L2}$,   the  $5\times 5$ flvaor mixing matrices  can be simplified as:
     \begin{align}
 V_\chi \approx  \left( \begin{array}{c|c}
  \mathbb{1}_{3\times 3} & - \pmb{ \epsilon}_\chi \\ \hline
 \pmb{ \epsilon}^\dagger_\chi  &  \mathbb{1}_{2\times 2}
 \end{array} \right)_{5\times 5}\,, \label{eq:VI_II}
  \end{align}
  where we only retain the leading contributions, and  the effects, which are smaller than $\pmb{\epsilon_\chi}$  with $\chi=R,L$, have been dropped, such as $\pmb{\epsilon^\dagger_\chi \epsilon_\chi}$,   $\pmb{m}_{\ell} \pmb{\delta m}_{1,2}/\pmb{m}^2_{L}$, etc.  The explicit expressions of $\pmb{\epsilon_\chi}$  then are  given by:
 \begin{align}
\pmb{ \epsilon^\dagger_L} & = \left(\begin{array}{ccc}
 0 \,, & \frac{v y_\mu }{\sqrt{2} m_{L1} }c_\alpha\,, &  - \frac{v y'_\tau}{\sqrt{2} m_{L1}} s_\alpha \\
  0\,, &   \frac{v y_\mu }{\sqrt{2} m_{L2}} s_\alpha\,, & \frac{v y'_\tau}{\sqrt{2} m_{L2}} c_\alpha
 \end{array} \right)  \,, ~~ 
 \pmb{\epsilon^\dagger_R } = \left(\begin{array}{ccc}
 0 \,, & -\frac{v_S y'_\mu }{\sqrt{2} m_{L1} }s_\alpha\,, &  \frac{v_S y_\tau}{\sqrt{2} m_{L1}} c_\alpha \\
  0\,, &   \frac{v_S y'_\mu }{\sqrt{2} m_{L2}} c_\alpha\,, & \frac{v_S y_\tau}{\sqrt{2} m_{L2}} s_\alpha
 \end{array} \right)  \,
 \end{align}
 where the Yukawa couplings $y_{\mu, \tau}$ and $y'_{\mu, \tau}$ are taken as  real numbers.

 After rotating the lepton weak states to physical states based on the $V_{R}$ and $V_L$, the Yukawa couplings of the SM Higgs and $\phi_S$ to the charged leptons from Eq.~(\ref{eq:Yukawa}) are expressed as:
 \begin{align}
 -{\cal L}_{h,\, \phi_S} &= \left(\begin{array}{cc}
    \pmb{\bar \ell_L }\,, & \pmb{\bar \Psi}_{\tau' L} \end{array} \right) V_L \left( \begin{array}{c|c}
  ~~~\pmb{ m_{\ell} }_{3\times 3}~~~ &  \pmb{ \delta m_1 } \\ \hline
0 &0
 \end{array} \right) V^\dagger_R
  \left(\begin{array}{c}
   \pmb{\ell_R} \\
      \pmb{\Psi}_{\tau' R} \end{array}\right) \frac{h}{v} \nonumber \\
    &+  \left(\begin{array}{cc}
  \pmb{\bar \ell_L }\,, & \pmb{\bar \Psi}_{\tau' L}  \end{array} \right)  V_L  \left( \begin{array}{c|c}
  ~~~0~~~ & 0  \\ \hline
\pmb{\delta m^T_2} &0
 \end{array} \right)  V^\dagger_R
  \left(\begin{array}{c}
   \pmb{\ell_R} \\
   \pmb{\Psi}_{\tau' R} \end{array}\right) \frac{\phi_S}{v_S}\,, \label{eq:LhS}
 \end{align}
 where we still use $\pmb{\ell}$ to represent the light leptons; however, for the mass eigenstate of the heavy lepton, we use $\pmb{\Psi}^T_{\tau'} =( \tau',\, \tau'') $ instead of $\pmb{\Psi}^T_{\ell'}=(\ell'_4,\, \ell'_5)$.  
 With the leading expansions in Eq.~(\ref{eq:VI_II}),  the Higgs and $\phi_S$ Yukawa couplings to the light charged leptons can be summarized as: 
 \begin{align}
-{\cal L}^{h, \phi_S}_Y & \supset \frac{m_\ell}{v} \bar \ell_L \ell_R h  
 -   \frac{ y_\mu y_\tau}{2 } \left( \frac{c^2_\alpha }{m_{L1}} + \frac{s^2_\alpha}{m_{L2}} \right) \bar \mu_L \tau_R ( v_S h + v \phi_S)  \nonumber \\
& -   \frac{ y'_\mu y'_\tau}{2}   \left( \frac{s^2_\alpha }{m_{L1}} + \frac{c^2_\alpha}{m_{L2}} \right)   \bar \tau_L \mu_R ( v_S h + v \phi_S ) + H.c. \label{eq:Yu_h_S}
 \end{align}
 The full Yukawa couplings are shown in the appendix; here, we  only  show the relevant parts. 
 From Eq.~(\ref{eq:L_h}), it can be seen that the modified Higgs couplings to $\mu$- and $\tau$-lepton are  proportional to $s_\alpha$ and $m_{L2}-m_{L1}$; thus, the modifications can be suppressed by a small $s_\alpha$ or/and $m_{L2}\approx m_{L1}$. Since the mixing between $\tau'$ and $\tau''$ does not influence the LFV effects,  in order to simplify the analysis, hereafter, we take $s_\alpha = 0$ and $c_\alpha = 1$.  According to Eq.~(\ref{eq:LhS}),  in addition to the $h\to  \mu \tau$ and $\tau \to \mu Z' Z'$ decays, the $h$- and $\phi_S$-mediated LFV effects can  also lead to a sizable muon $g-2$. Since the LFV currents involve $\bar \tau_L \mu_R$ and $\bar\mu_L \tau_R$, the muon $g-2$ can be enhanced by $m_\tau$ due to  the $\tau$ chirality flip.  
  Although the  $\tau \to 3 \mu$ decay is allowed in the model, since the vertices are suppressed by the $m_\mu/v$  and have no other enhancing factor, the resulting  branching ratio is $\sim 8 \times 10^{-13}$  and is far below the current experimental upper bound of $2.1 \times 10^{-8}$~\cite{PDG}. Similarly, the BR for the $\tau\to \mu \gamma$ decay is also small. The  LFV couplings, which involve heavy leptons, can be found in the appendix.

Next, we discuss the influence of the vector-like leptons on the $Z$ and $Z'$ gauge couplings  to the leptons. Since the introduced heavy leptons are $SU(2)_L$ singlets and carry the hypercharge $Y=-1$, which is the same as that carried by the right-handed light leptons,  the  $Z$ gauge couplings to the right-handed leptons are flavor conserving at the tree level. However, because the left-handed light and heavy leptons carry different $U(1)_Y$ charges, the Z-mediated LFCNCs at the tree level occur in the left-handed leptons. To show the newly modified $Z$ gauge couplings to the leptons, we write the interactions in the physical lepton states as:
 \begin{align}
 {\cal L}_Z &= - \left( \pmb{\bar \ell}_L ,  \pmb{\bar \Psi}_{ \tau' L}  \right) \gamma_\mu
      V_L \left( \begin{array}{c|c}
  ~~~C^\ell_L \mathbb{1}_{3\times 3}~~~ &  0 \\ \hline
0 & C^\ell_R  \mathbb{1}_{2\times 2}
 \end{array} \right) V^\dagger_L
  \left(\begin{array}{c}
   \pmb{\ell_L} \\
     \pmb{\Psi}_{ \tau' L} \end{array}\right) Z^\mu\,, \\
  &  \approx - \left( \pmb{\bar \ell}_L , \pmb{\bar \Psi}_{\tau' L}\right) \gamma_\mu
       \left( \begin{array}{c|c}
  C^\ell_L \mathbb{1}_{3\times 3} &  (C^\ell_L - C^\ell_R) \pmb{\epsilon}_L \\ \hline
 (C^\ell_L - C^\ell_R) \pmb{\epsilon}^\dagger_L & C^\ell_R  \mathbb{1}_{2\times 2}
 \end{array} \right) 
  \left(\begin{array}{c}
   \pmb{\ell_L} \\ 
   \pmb{\Psi}_{ \tau' L} \end{array}\right) Z^\mu\,, \\
    C^\ell_L & = \frac{g}{2c_W} \left( 2s^2_W -1 \right)\,, \quad C^\ell_R = \frac{g s^2_W}{c_W}\,,
 \end{align}
where we have dropped the $\pmb{\epsilon}_L \pmb{\epsilon}^\dagger_L$ and $\pmb{\epsilon}^\dagger_L \pmb{\epsilon}_L$ effects; $c_W =\cos\theta_W$, $s_W=\sin\theta_W$, and $\theta_W$ is the Weinberg's angle.  Based on this approximation, the $Z$-boson couplings to the light charged leptons are still flavor conserving.  Nevertheless, the LFV processes can occur in the transition of the heavy leptons to  light leptons. Hence, with the leading order approximation to the flavor mixing matrices, the $Z$-boson couplings to the charged leptons  can be simplified as:
  \begin{align}
  -{\cal L}_{Z} & \approx \left[ C^\ell_L \pmb{\bar \ell}_L \gamma_\mu  \pmb{\ell}_L +C^\ell_R  \pmb{\bar \ell}_R \gamma_\mu  \pmb{\ell}_R+ C^\ell_R \pmb{\bar \Psi}_{\tau'} \gamma_\mu  \pmb{ \Psi}_{\tau'}  \right] Z^\mu  \nonumber \\
   &- \frac{g}{2c_W} \left( \frac{v y_\mu}{\sqrt{2} m_{L1} } \bar\mu_L \gamma_\mu  \tau'_L + \frac{ v y'_\tau}{\sqrt{2} m_{L2}} \bar\tau_L \gamma_\mu \tau''_L  + H.c. \right) Z^\mu \,. \label{eq:ZI}
  \end{align}

Although the introduced $\ell_{4,5}$ leptons are vectorial couplings to the $Z'$ boson, since the charged leptons carry different $U(1)_{\mu-\tau}$ charges, and the flavor mixing matrices in general distinguish the lepton chirality,  we thus write 
 the $Z'$ couplings to the leptons as:
\begin{align}
 {\cal L}_{Z'} & = -g_{Z'} \bar \ell' \gamma^\mu V_R Q' V^\dagger_R  P_R \ell' Z'_\mu - g_{Z'} \bar \ell' \gamma^\mu V_L Q' V^\dagger_L  P_L \ell' Z'_\mu - g_{Z'} \bar \nu_\ell \gamma^\mu Q P_L \nu_{\ell} Z'_\mu\,,
 \end{align}
where $\ell'^T=( e, \mu, \tau, \tau', \tau'')$; dia$Q'=(0,1,-1, 1, -1)$ and  dia$Q=(0, 1, -1)$  denote the $U(1)'$ charges of the charged leptons and neutrinos, respectively.  Taking the leading approximation, the couplings to the charged leptons can be simplified as:
 \begin{align}
 - {\cal L}_{Z'} &\approx g_{Z'}    \left( \begin{array} {cc}
  \pmb{\bar \ell}, &  \pmb{\bar \Psi}_{\tau'} \end{array}\right)  \gamma^\mu  Q'  
 \left( \begin{array}{c}
   \pmb{\ell} \\
    \pmb{\Psi}_{\tau'} \end{array} \right) Z'_\mu   \nonumber \\
    &-  g_{Z'}  \frac{2 v_S y_\tau }{\sqrt{2} m_{L1}} \bar \tau_R \gamma^\mu   \tau'_R Z'_\mu   + g_{Z'}  \frac{2 v_S y'_\mu }{\sqrt{2} m_{L2}} \bar \mu_R \gamma^\mu   \tau''_R Z'_\mu + H.c. 
 \end{align}
It can be seen that the $Z'$ couplings to the lepton pairs are still flavor conserved. The $Z'$ mediated flavor changing effects only occur in the right-handed currents and at the $\tau'-\tau$ and $\tau''-\mu$ vertices. Since we focus on the light $Z'$ and small $g_{Z'}$, these flavor-changing couplings cannot have a significance influence on the muon $g-2$. In addition, the contributions to the $\tau'$ and $\tau''$ decay widths are also small.

\section{Numerical Analysis and Discussions}

Based on the introduced interactions, in the following, we discuss the relevant phenomena of interest. 

\subsection{$h\to \mu \tau$ and muon $g-2$}

\begin{figure}[t]
\includegraphics[width=85mm]{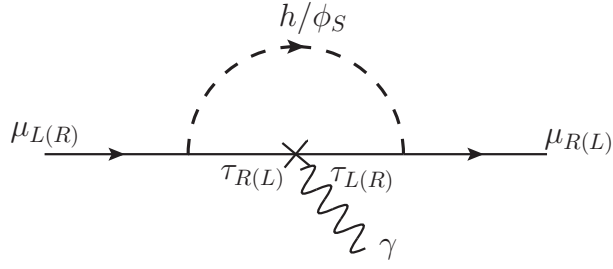}  
\caption{ Sketched Feynman diagram for the Higgs- and $\phi_S$-mediated muon $g-2$.}
\label{fig:fey_gm2}
\end{figure}

\begin{figure}[hptb]
 \includegraphics[width=75mm]{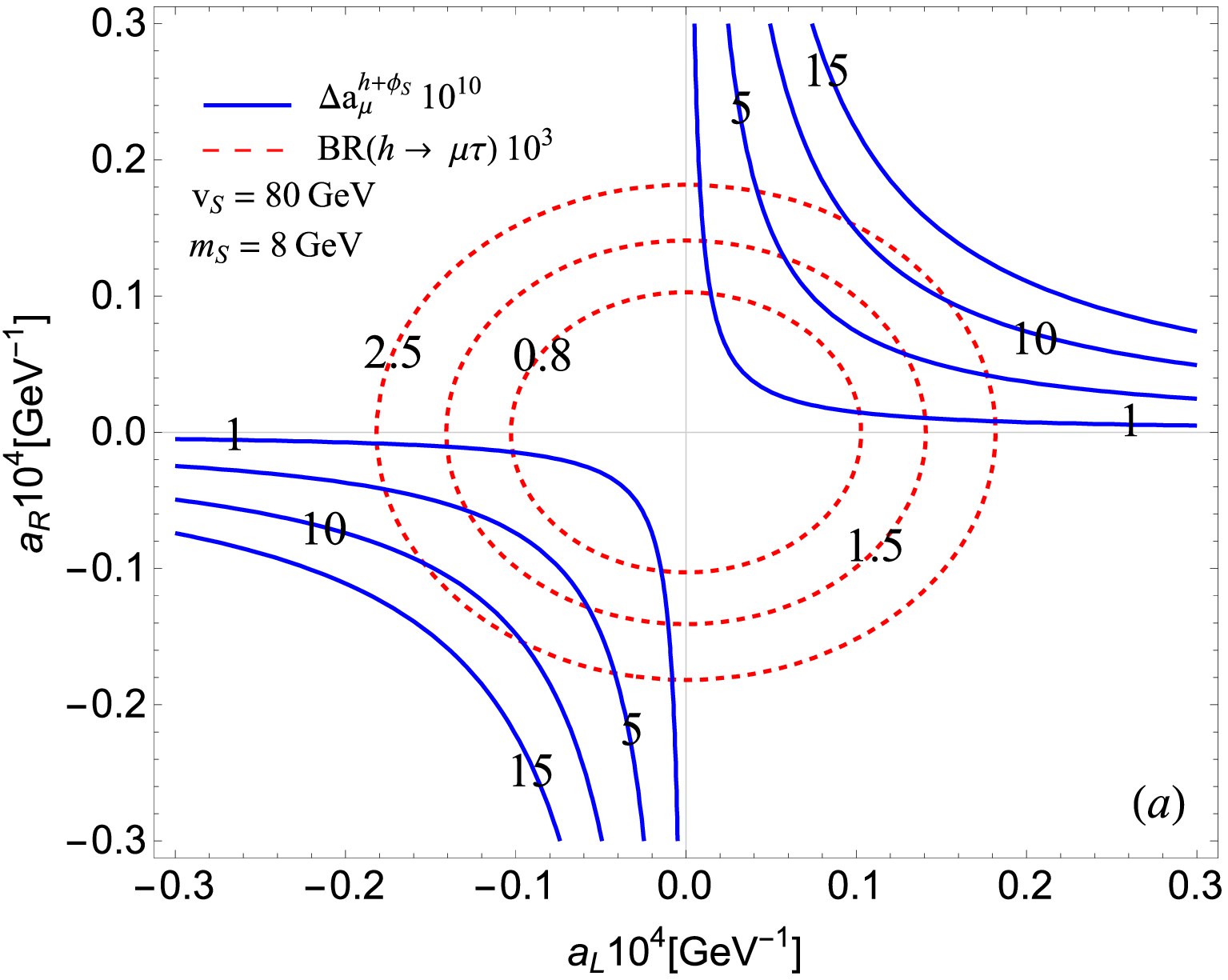}  
\includegraphics[width=75mm]{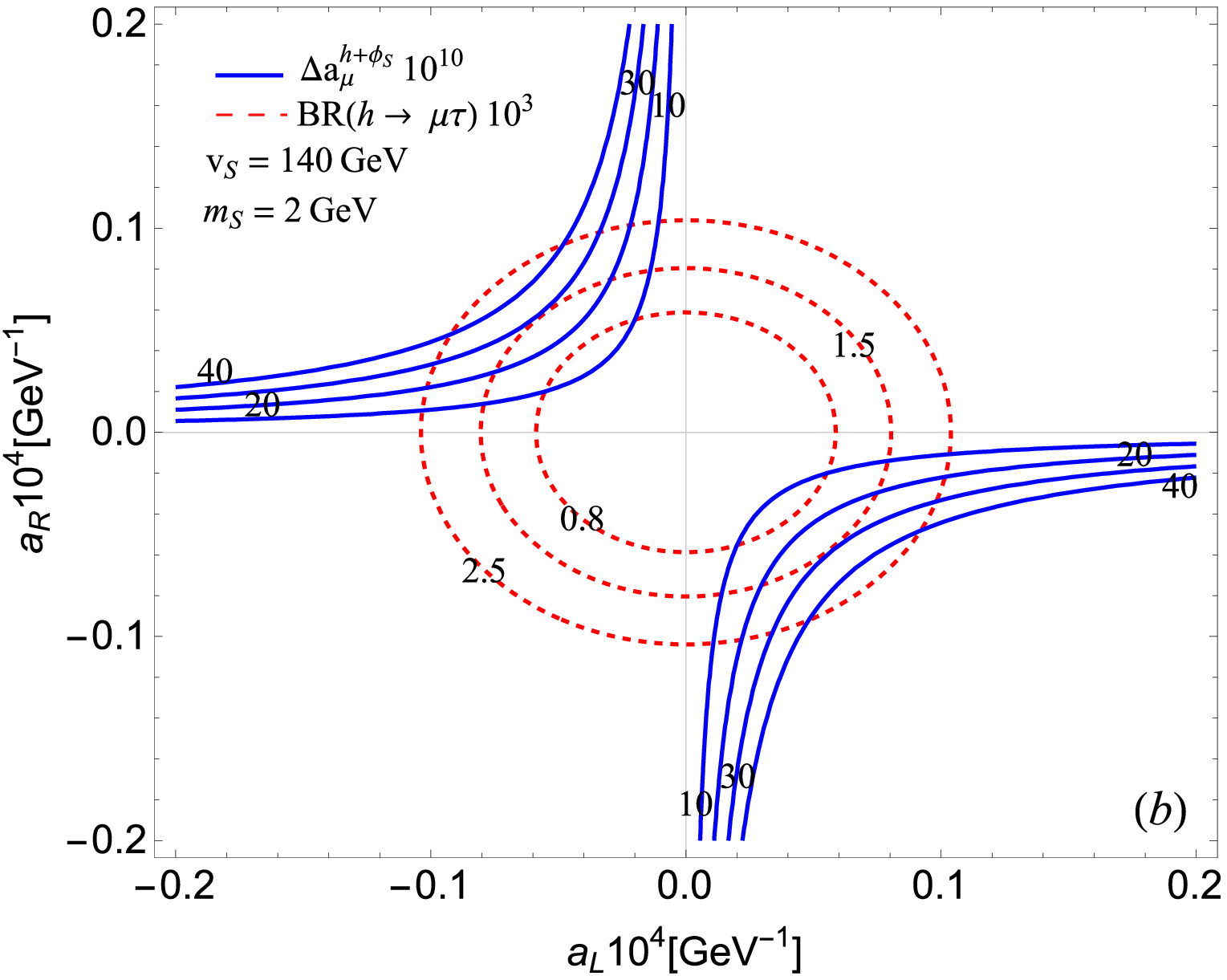} 
\caption{   Contours for $BR(h\to  \mu \tau)$ in units of $10^{-3}$ (dashed) and  $\Delta a_\mu^{h+\phi_S}$ in units of $10^{-10}$ (solid)   as a function of $a_R$ and $a_L$, where  $(v_S, m_S)=(80, 8)$ GeV  in plot (a) and $(v_S, m_S)=(140, 2)$ GeV in plot (b) are taken.  }
\label{fig:g-2}
\end{figure}

From Eq.~(\ref{eq:Yu_h_S}), it is found that the  LFV coupling  $\mu-\tau-h$ leads to the $h \to  \mu \tau$ decay, and  the associated BR can be written  as:
\begin{align}
BR(h\to  \mu \tau) &= \frac{ v^2_S \left(|a_L|^2 + |a_R|^2\right)}{8 \pi  \Gamma_h} m_h  \label{eq:brhtaumu}\,,\\
%
%
 a_L & = \frac{ y_\mu y_\tau}{2 m_{L1}}\,, \quad a_R = \frac{ y'_\mu y'_\tau }{2 m_{L2}} \,, \nonumber
\end{align}
where $\tau \mu$ indicates the sum of $\bar \mu \tau + \bar\tau \mu$, and $\Gamma_{h}$ is the Higgs width.  
With $m_h=125$ GeV, $\Gamma_h\approx 4.21$ MeV,  the parameters can be reformulated as:
 \begin{equation}
  \sqrt{|a_L|^2 + |a_R|^2} \approx \frac{ 1.56\times 10^{-3}}{v_S} \sqrt{\frac{BR(h\to  \tau \mu)}{2.5\times 10^{-3}}}\,, \label{eq:aLR}
 \end{equation}
where $BR(h\to  \mu \tau)$ can be taken from the experimental data, and  the current upper limits from ATLAS and CMS are $1.43\%$~\cite{Aad:2016blu} and $0.25\%$ ~\cite{Khachatryan:2015kon,CMS:2017onh}, respectively. Hereafter, we take $2.5\times 10^{-3}$ as the upper limit of $BR(h\to \mu\tau)$.  Moreover, the LFV effects in Eq.~(\ref{eq:Yu_h_S}) can also contribute to the muon $g-2$, for which the  Feynman diagram is sketched  in Fig.~\ref{fig:fey_gm2}. 
Thus, in addition to the $Z'$-mediated loop effect, the new sources contributing to the muon $g-2$ in this model are from the $h$- and $\phi_S$-mediated  loop diagrams. 
As a result,  the muon $g-2$ can be written as:
\begin{equation}
\Delta a_\mu = \Delta a_\mu^{Z'} + \Delta a_\mu^{h} + \Delta a_\mu^{\phi_S},
\label{eq:amu_total} 
\end{equation}
where the current measurement is $\Delta a_\mu =a^{\rm exp}_\mu - a^{\rm SM}_\mu= (28.7 \pm 8.0)\times 10^{-10}$~\cite{PDG},  the $Z'$ contribution is given as~\cite{Araki:2017wyg}:
  \begin{equation}
  \Delta a^{Z'}_\mu = \frac{g^2_{Z'}}{8 \pi^2 } \int^1_0 dx \frac{2 m^2_\mu x^2 (1-x)}{ x^2 m^2_\mu + (1-x) m^2_{Z'} }\,, 
  \label{eq:amu_Zprime}
  \end{equation}
and the $h$ and $\phi_S$ effects are respectively expressed as:
 \begin{align}
  \Delta a_\mu^h & = -Q_\tau m_\mu m_\tau \frac{  a_R a_L}{4\pi^2 }  \frac{v^2_S }{m^2_h} \left( \ln\frac{m^2_h}{m^2_\tau} -\frac{3}{2}\right) \,, \nonumber  \\ 
 \Delta a_\mu^{\phi_S} & = -Q_\tau m_\mu m_\tau \frac{  a_R a_L}{4\pi^2 }  \frac{v^2 }{m^2_S} \left( \ln\frac{m^2_S}{m^2_\tau} -\frac{3}{2}\right)  
 \label{eq:Da_mu}
 \end{align}
 with $Q_\tau=-1$ being the $\tau$-lepton electric charge.  From Eqs.~(\ref{eq:brhtaumu}) and (\ref{eq:Da_mu}), it can be seen that the scalar contributions to the $h\to \mu \tau$ decay are dictated by $|a_L|^2+|a_R|^2$ while the contributions to the  muon $g-2$,  denoted by  $\Delta a_\mu^{h+\phi_S} \equiv \Delta a_\mu^h + \Delta a_\mu^{\phi_S}$, are associated with $a_L a_R$. When one of $a_{L}$ and $a_{R}$ is small or vanishes, $BR(h\to \mu \tau)$ can still be sizable, however, $\Delta a_\mu^{h+\phi_S}$ is suppressed.  
 In order to investigate the case when the  $BR(h\to \mu \tau)$ and $\Delta a_\mu^{h+\phi_S}$ are strongly correlated, in the following analysis, we take the scheme with $|a_L|\approx |a_R|$. 
 In addition, since the sign of $\zeta=\ln(m^2_S/m^2_\tau) - 3/2$ depends on $m_S$,   in order to get a positive  $\Delta a_\mu^{h+\phi_S}$,  the relative sign of $a_R$ and $a_L$ also depends on the value of $m_S$. We show the contours for the $BR(h\to \mu \tau)$ in units of $10^{-3}$ (dashed) and  $\Delta a_\mu^{h+\phi_S}$ in units of $10^{-10}$ (solid) as a function of $a_R$ and $a_L$ in Fig.~\ref{fig:g-2}(a), where $v_S=80$ GeV and $m_S =8$ GeV are used.   Similarly, we show the case with $v_S=140$ GeV and $m_S=2$  GeV  in Fig.~\ref{fig:g-2}(b).  From the plots, it can be found that  $BR(h\to \mu \tau)$ of $10^{-3}$ and $\Delta a_\mu^{h+\phi_S}$ of $10\times 10^{-10}$ can be reconciled by the scalar-mediated LFV effects; and $a_L$ and $a_R$ prefer the same sign in plot (a) while they are  opposite sign in plot (b). 

 \begin{figure}[hptb]
\includegraphics[width=75mm]{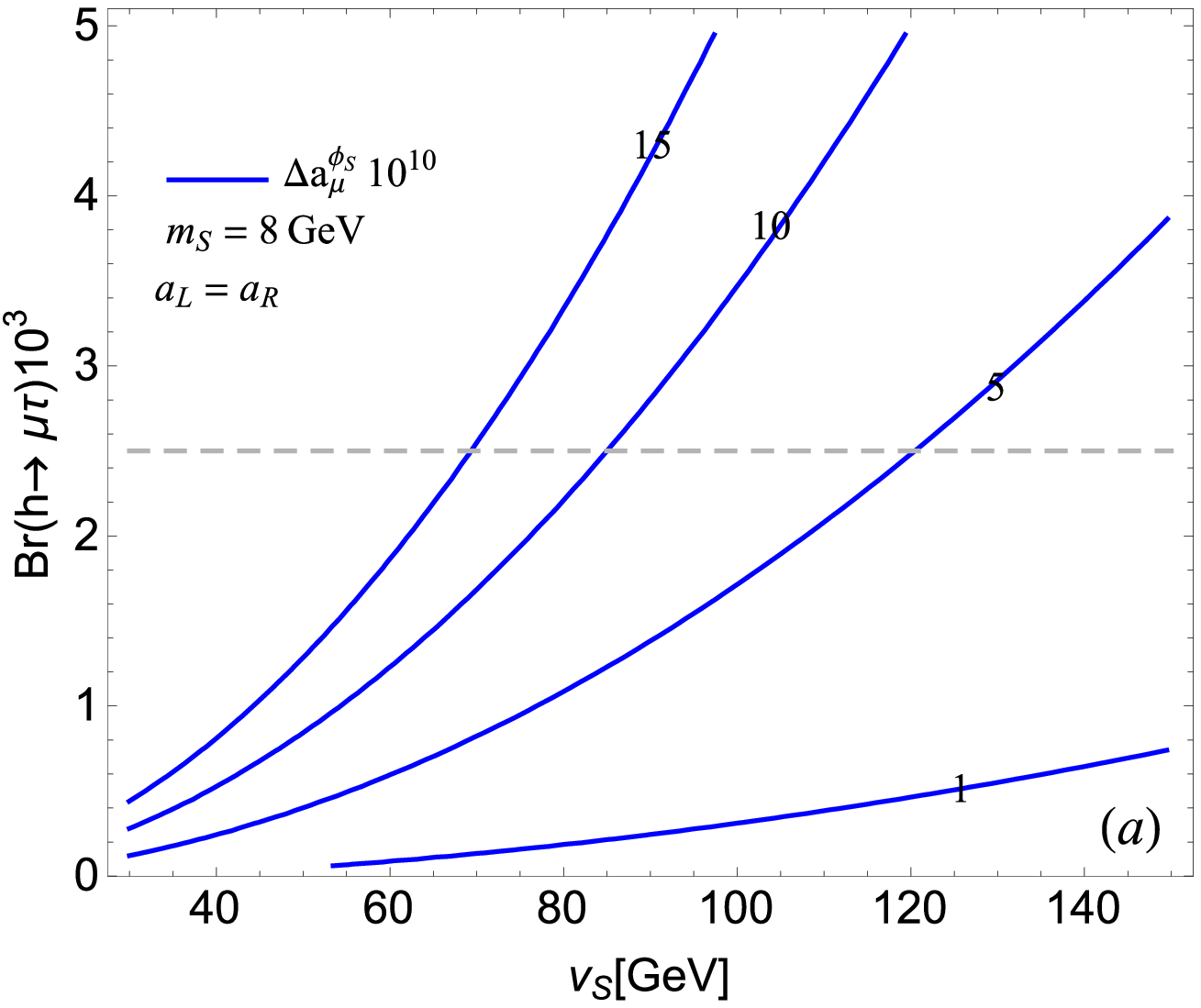}  
\includegraphics[width=75mm]{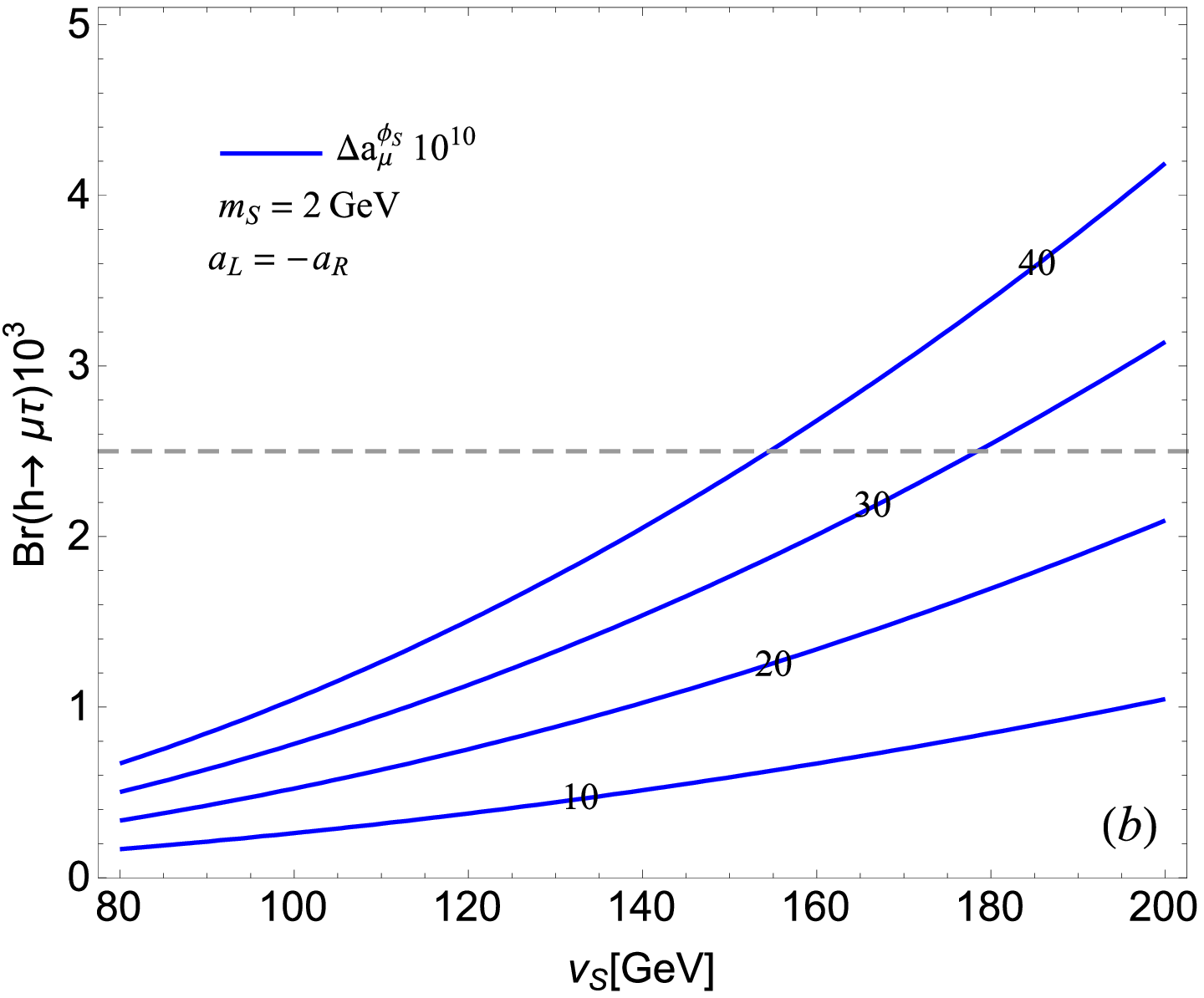} 
\caption{   Contours for  $\Delta a_\mu^{\phi_S}$ (in units of $10^{-10}$)  as a function of $BR(h\to  \mu \tau)$ and $v_S$, where $m_S =8$ GeV and  $a_L= a_R$ are used in plot (a), $m_S =2$ GeV and  $a_L= -a_R$ are used in plot (b), and the dashed line is the upper limit from CMS~\cite{CMS:2017onh}.}
\label{fig:g-2brh}
\end{figure}

 Basically,  $a_R$ and $a_L$ appearing in Eqs.~(\ref{eq:brhtaumu}) and (\ref{eq:Da_mu}) can be taken as  two independent parameters; however,  if we use the scheme with $a_L \approx \pm a_R$,  $\Delta a_\mu^{h+\phi_S}$  can be expressed in terms of  $BR(h\to \mu \tau)$ as:
 \begin{align}
  \frac{ \Delta a_\mu^{h+\phi_S}}{BR(h\to \mu \tau) } \approx \pm \frac{m_\mu m_\tau \Gamma_h}{\pi m^3_h} \left[ \left( \ln\frac{m^2_h}{m^2_\tau} -\frac{3}{2}\right) + \frac{ (v m_h)^2 }{(m_S v_S)^2} \left( \ln\frac{m^2_S}{m^2_\tau} -\frac{3}{2}\right) \right]\,.\label{eq:g-2_b}
  \end{align}
 Since there is no other free parameter in the first term of Eq.~(\ref{eq:g-2_b}),  if we take the CMS upper limit with  $BR(h\to  \mu \tau) \sim 2.5\times 10^{-3}$, the Higgs contribution   can be estimated to be $\Delta a^h_\mu \sim 2.2\times 10^{-12}$, which is far below the current experimental value. Hence, the dominant contribution to  $\Delta a_\mu^{h+\phi_S}$ is from the $\phi_S$ mediation. For simplicity,  we take $\Delta a_\mu^{h+\phi_S} \simeq \Delta a_\mu^{\phi_S}$ in our analysis.   Based on Eq.~(\ref{eq:g-2_b}), we show the contours for  $\Delta a_\mu^{\phi_S}$ (in units of $10^{-10})$ as a function of $BR(h\to  \mu \tau)$ and $v_S$ in Fig.~\ref{fig:g-2brh}(a) and  Fig.~\ref{fig:g-2brh}(b), where  the former plot corresponds to $m_S=8$ GeV and $a_L=a_R$,  the latter plot is $m_S=2$ GeV and $a_L=-a_R$, and the dashed line in both plots denotes the CMS upper limit.  From the plots, it can be seen that when the value of $BR(h\to  \mu \tau)$ is around $1\times 10^{-3}$, the value of  $\Delta a_\mu^{\phi_S}$ can still reach $10\times 10^{-10}$. 
 
\begin{figure}[phtb]
\includegraphics[width=75mm]{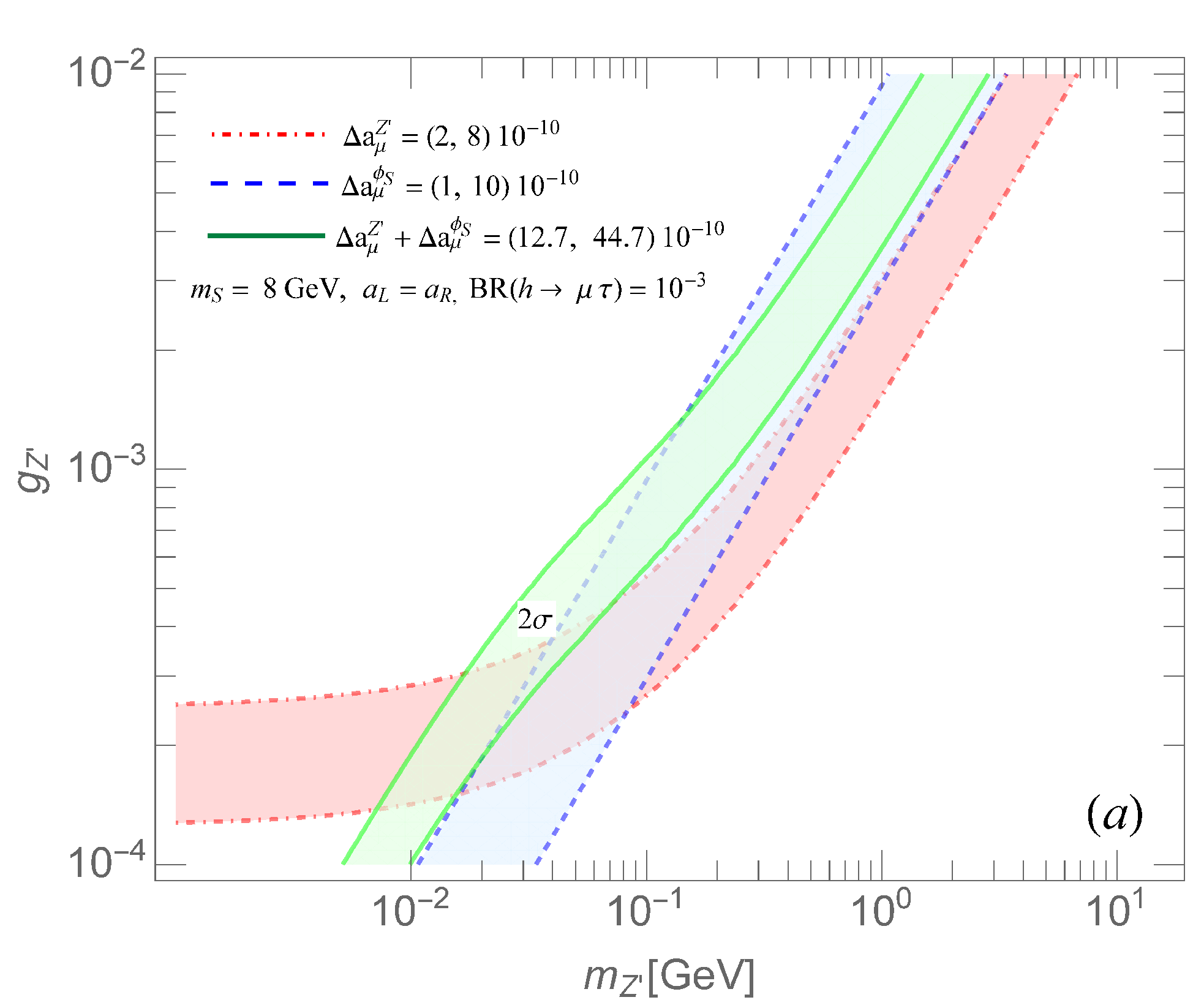}  
\includegraphics[width=75mm]{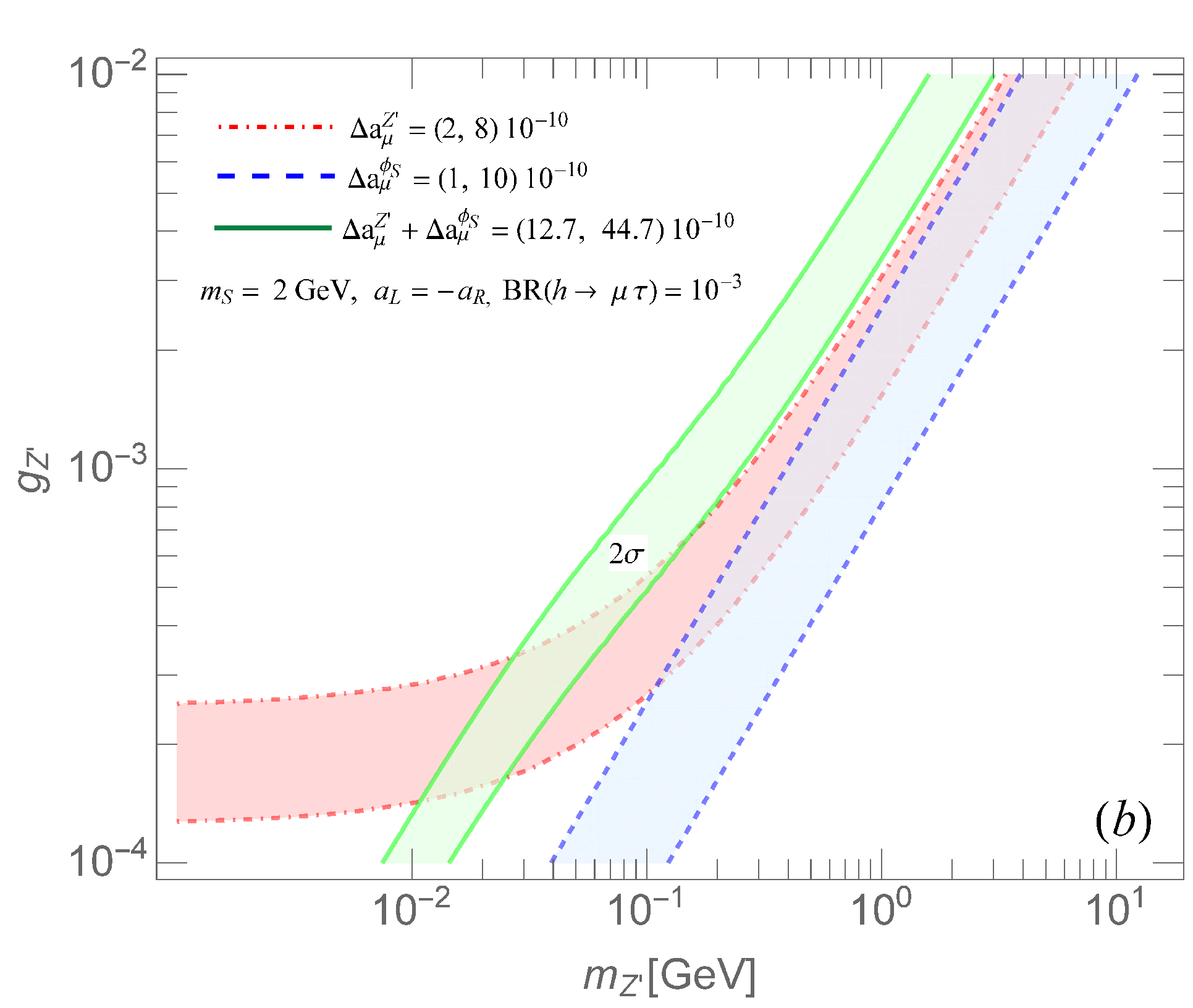} 
\caption{  Contours for $\Delta a_\mu$ with (a) $m_S=8$ GeV and $a_L=a_R$ and  (b) $m_S=2$ GeV and $a_L=-a_R$ as a function of $m_{Z'}$ and $g_{Z'}$, where the results bounded by the dot-dashed, dashed, and solid lines denote the $Z'$, $\phi_S$, and $Z'+\phi_S$ contributions, respectively.  }
\label{fig:gm2_phiS_Zp}
\end{figure}

  In order to clearly see the influence of the $\phi_S$-mediated effects on the muon $g-2$, we  show the contours for  $\Delta a_\mu^{\phi_S}$ as a function of $m_{Z'}$ and $g_{Z'}$ in Fig.~\ref{fig:gm2_phiS_Zp}, where $v_S= m_{Z'}/(2 g_{Z'})$,  $BR(h\to \mu \tau)=1\times 10^{-3}$, $m_S=8$ GeV in plot (a), and $m_S=2$ GeV in plot (b) are used; the results bounded by the dot-dashed, dashed, and solid lines represent the contributions from the $Z'$ boson, $\phi_S$, and $Z'+\phi_S$, respectively; the taken region for each contribution  is given by $\Delta a^{Z'}_{\mu} =(2, 8) \times 10^{-10}$, $\Delta a^{\phi_S}_\mu= (1, 10) \times 10^{-10}$, and $\Delta a^{Z'+\phi_S}_\mu = (12.7, 44.7)\times 10^{-10}$.
  ; and the $Z'$ contribution is written as Eq.~(\ref{eq:amu_Zprime}). 
  Since $\Delta a^{\phi_S}_\mu \propto  BR(h\to \mu \tau)/v^2_S = 4 g^2_{Z'}BR(h\to \mu \tau)/m^2_{Z'}$, when $m_{Z'}$ approaches the region of $m_{Z'}\ll 1$ GeV, $g_{Z'}$ must decrease in order to keep $\Delta a^{\phi_{S}}_\mu$ constant. This behavior is different from the $Z'$-mediated muon $g-2$, where $\Delta a^{Z'}_\mu$ approaches a constant when $m_{Z'}$ goes to zero.   According to the plots,  if  the $Z'$-mediated muon $g-2$ is $\Delta a^{Z'}_\mu < 10\times 10^{-10}$ at $g_{Z'}\sim {\cal O}(10^{-3})$,    the muon $g-2$ can be enhanced to  the current data with $2\sigma$ errors when   the $\phi_S$ contribution is included. 

\subsection{$Z'$ and rare $\tau$ decays}

\begin{figure}[phtb]
\includegraphics[width=75mm]{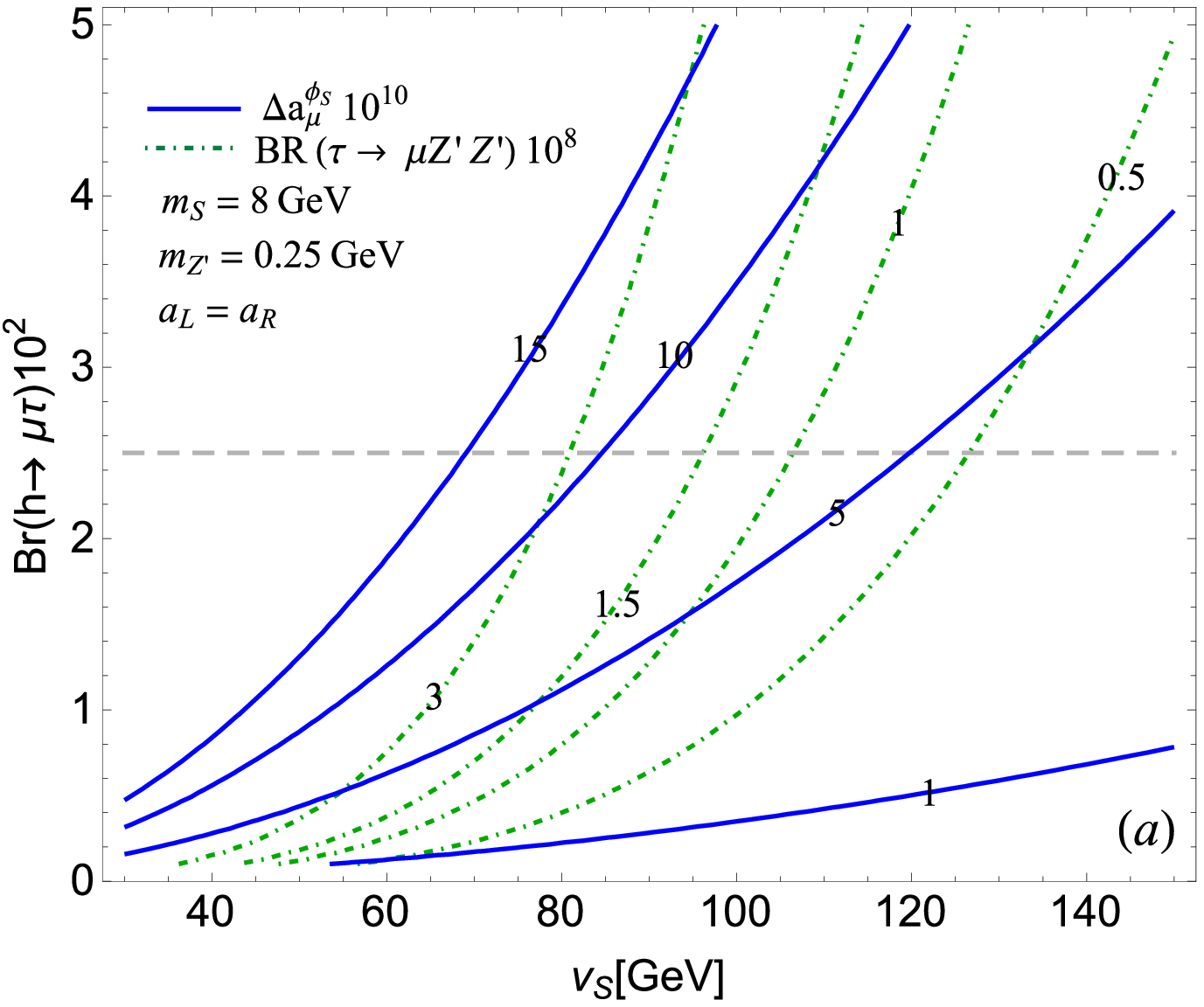}  
\includegraphics[width=75mm]{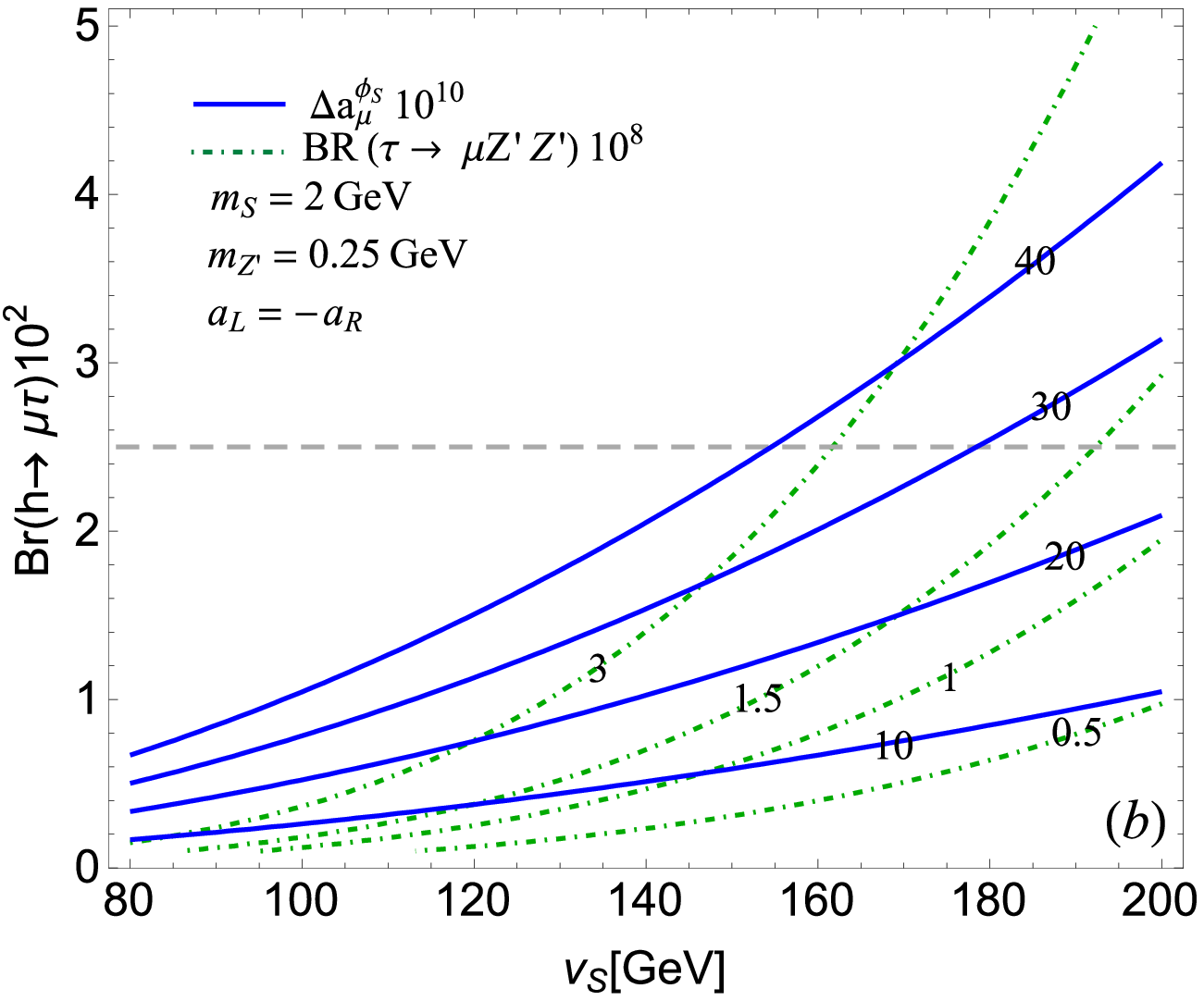}
\caption{  Contours for $BR(\tau\to \mu Z'Z')$ in units of $10^{-8}$ (dot-dashed) and  $\Delta a_\mu^{\phi_S}$ in units of $10^{-10}$ (solid)  with $m_{Z'}=0.25$ GeV as a function of  $BR(h\to  \mu \tau)$ and $v_S$, where plot (a)  denotes $m_S =8$ GeV and $a_L=a_R$, and the plot (b) is $m_S=2$ GeV and $a_L=-a_R$.  }
\label{fig:taumu2Zp}
\end{figure}

As shown above,  in order to enhance the muon $g-2$ up to the  $10^{-10}-10^{-9}$ level,    a light $\phi_S$ is preferred. In this situation, the $\phi_S$ predominantly decays into $\tau \mu$ and $Z' Z'$.  According to the gauge coupling in Eq.~(\ref{eq:SZ'Z'}) and the Yukawa couplings in Eq.~(\ref{eq:Yu_h_S}), the $\phi_S$ partial decay rates 
can be expressed as:
 \begin{align}
  \Gamma(\phi_S \to Z' Z') & =  \frac{m^3_S}{16\pi v^2_S}\,,  \nonumber \\
  \Gamma(\phi_S \to \tau \mu) & = \frac{v^2 \left( |a_L|^2 + |a_R|^2 \right)}{8 \pi }  m_{S}\,, \nonumber  \\
  & =  8.3 \times 10^{-8} \frac{m_S v^2 }{ v^2_S} \frac{BR(h\to\tau \mu)}{2.5\times 10^{-3}}\,, \label{eq:brS} 
 \end{align}
where the lepton and $Z'$ mass effects are neglected, and $\tau \mu$ indicates the sum of the $\bar\tau \mu$ and $\bar\mu \tau$ channels. In the last line, we applied the result of Eq.~(\ref{eq:aLR}).  It can be clearly seen that $\Gamma(\phi_S \to \tau \mu) \ll \Gamma(\phi_S \to Z'Z')$. Accordingly, the BRs for the $\phi_S$ decays are shown as:
 \begin{align}
 BR(\phi_S\to Z' Z')  & \approx 1 \,, \nonumber \\ 
 BR(\phi_S \to \tau \mu) & \approx 2.57 \times 10^{-3} \left( \frac{10\, \rm GeV}{ m_S}\right)^2  \frac{BR(h\to \mu \tau)}{2.5\times 10^{-3}}\,.
 \end{align}  
 In addition to the $\phi_S \to \tau \mu$ decay, the same LFV effects can lead to $\tau\to  \mu  Z'Z'$ decay through the $\phi_S$ mediation.  The differential branching fraction as a function of $Z'Z'$ invariant mass is shown as:
 \begin{align}
 \frac{dBR(\tau \to \mu Z'  Z' )}{dq^2} &  \approx  \frac{m_\tau}{64\pi^2 m_h} \frac{\Gamma_h}{ \Gamma_\tau } BR(h\to \mu \tau) \nonumber \\
  & \times  \frac{ (q^2 -2 m_{Z'}^2)^2 + 8 m^4_{Z'}}{v^4_S m^2_S}\left( 1- \frac{q^2}{m^2_{\tau}} \right)^2 \sqrt{1- \frac{4 m^2_{Z'}}{q^2}}\,.
 \end{align}
 Based on the result, we show the contours for $BR(\tau \to \mu Z'Z')$ (dot-dashed) as a function of $BR(h\to \mu \tau)$ and $v_S$ with $m_{Z'}=0.25$ GeV in Fig.~\ref{fig:taumu2Zp}, where plot (a) denotes $m_S=8$ GeV and $a_L=a_R$, and plot (b) is $m_S=2$ GeV and $a_L=-a_R$. For comparison, we also show the muon $g-2$ (solid) in the  plot (a) and (b), and the numbers on the contour lines denote the values of $BR(\tau \to \mu Z'Z')$ and  $\Delta a_\mu^{\phi_S}$, which have been rescaled by a $10^{-8}$ and a $10^{-10}$ factor, respectively. From the plots, it can be seen that when  $\Delta a_\mu^{\phi_S}$ is of the order of $10^{-10}$, the associated $BR(\tau \to \mu Z'Z')$ is in the order of $10^{-8}$. 

The possible detecting signals for $\tau \to \mu Z'Z'$  depend on the $Z'$ mass. If $m_{Z'}< 2 m_{\mu}$, the $Z'$ gauge boson can only decay into $\nu_\mu$ and $\nu_\tau$ pairs. Thus, the detecting signals will be $\tau \to \mu + \slashed{E}$ with $\slashed{E}$ being a missing energy. Since the $\tau \to \mu  \bar\nu_\nu \nu_\tau$ $BR\approx 17.27\%$ process in the SM is the main background, the small $BR(\tau \to \mu Z'Z')$ cannot be distinguished from the $BR^{\rm exp}(\tau \to \mu \bar\nu_\nu \nu_\tau)=(17.39 \pm 0.04)\%$ errors~\cite{PDG}. In this case, we cannot see the signal for the $\tau\to \mu Z'Z'$ decay.  However, when $m_{Z'}> 2 m_{\mu}$, in addition to the neutrino pair, the $Z'$ can also decay into a muon pair. Therefore, the signals can be
$\tau \to 3 \mu \, \bar \nu\, \nu$ and $\tau \to 5 \mu$, where $\nu$ includes the $\nu_\mu$ and $\nu_\tau$ neutrinos, and the decay chains are shown as:
 \begin{align}
& \tau \to \mu Z'Z';  Z' \to \bar \nu \nu,\, Z' \to \mu^+ \mu^-  \\
 & \tau \to \mu Z'Z';  Z' \to \mu^+ \mu^-,\, Z' \to \mu^+ \mu^-\,.
 \end{align}
The BRs for the $Z'\to (\bar \nu \nu, \mu^+ \mu^-)$ decays can be simply formulated as:
\begin{align}
BR(Z'\to \bar\nu \nu) &=  \frac{1}{1+M^2_\mu} \,, \quad BR(Z'\to \mu^+\mu^-)=\frac{M^2_\mu}{ 1 + M^2_\mu}\,,\\
M^2_\mu & = \left( 1 + \frac{2 m^2_\mu}{m^2_{Z'}}\right)\sqrt{1-\frac{4m^2_\mu}{m^2_{Z'}}}\,,
\end{align}
where  the BRs only depend on the $m_{Z'}$ parameter.
For illustration, we show the values of BRs with respect to some selected $m_{Z'}$ values in Table~\ref{tab:BRZp}. According to the results in Fig.~\ref{fig:taumu2Zp} and the values shown in Table~\ref{tab:BRZp}, it can be found that the BRs for $\tau\to (3\mu+\slashed{E}, 5\mu)$ can reach a level of $10^{-9}$, which is the detecting sensitivity at the Belle II~\cite{Flores-Tlalpa:2015vga}. 
 In the Bell II experiment, $\tau$ leptons are produced in pairs and the signals of the $\tau$ decays could be $e^+ e^- \to \tau (\to 3 \mu + \slashed{E}, 5 \mu) + \tau_h$, where $\tau_h$ is  the hadronic $\tau$ decays. Thus, the SM background events are $(3 \mu + \slashed{E}, 5 \mu) +$ jet, which can be produced by the  electroweak interactions. The background events can be reduced by applying proper kinematic cuts, such as  the $\tau$ mass reconstruction  from $3 \mu + \slashed{E}$ or  $5 \mu$ in the final state, the muon-pair invariant mass distribution, and the various kinematic distributions of the same sign muons.

\begin{table}[htp]
\caption{ Values of branching ratios for the $Z'\to (\bar \nu \nu, \mu^+ \mu^-)$ decays with respect to the selected $m_{Z'}$ values, where $\nu$ includes the $\nu_\mu$ and $\nu_\tau$ neutrinos. }
\begin{tabular}{cccccc} \hline \hline
      ~~~$m_{Z'}$  [GeV] ~~~   &  ~~~$0.22$~~~  & ~~~$0.25$~~~ & ~~~ $0.3$ ~~~& ~~~ $0.34$  ~~~ & ~~~ $0.38$  ~~~\\ \hline 
 $BR(Z'\to \bar \nu \nu)$ &   0.71   &   0.58      &  0.53      &  0.52  & 0.51 \\ \hline 
 $BR(Z'\to \mu^+ \mu^-)$ &  0.29   &   0.42       & 0.47      &  0.48  & 0.49  \\\hline\hline

\end{tabular}
\label{tab:BRZp}
\end{table}%

\subsection{ Influence on $e^+ e^- \to \gamma Z'$}

Next, we examine the influence of vector-like leptons on the $e^+ e^- \to \gamma Z'$ process, which arises from the kinetic mixing~\cite{Araki:2017wyg}. 
The Feynman diagram for the loop-induced kinetic mixing is sketched in Fig.~(\ref{fig:km}), where the leptons inside the loop include $\ell'=\mu, \tau, \tau'$, and $\tau''$. Accordingly, the effective Lagrangian is expressed as:
 \begin{align}
 {\cal L}_{\rm mix}& = -\frac{\epsilon}{2}  F_{\mu \nu} Z'^{\mu \nu} \,, \nonumber \\
  &= - \Pi(q^2) \left( q^2 \epsilon_\gamma \cdot \epsilon^*_{Z'} - q\cdot \epsilon_\gamma q\cdot \epsilon^*_{Z'} \right)\,,
 \end{align}
where $F_{\mu\nu}$ and $Z'_{\mu \nu}$ are the $U(1)_{\rm em}$ and $U(1)_{\mu-\tau}$ gauge field strength tensors, respectively, and the $\epsilon=\Pi(q^2)$ can be derived as:
 \begin{equation}
 \Pi(q^2)  = \frac{8 e g_{Z'}}{(4\pi)^2} \int^{1}_{0} dx\, x(1-x)\left[ \ln\frac{m^2_\tau - x(1-x) q^2}{m^2_\mu - x(1-x) q^2} + \ln\frac{m^2_{L2} - x(1-x) q^2}{m^2_{L1} - x(1-x) q^2} \right]\,. \label{eq:Pi}
 \end{equation}
Since the $U(1)_{\mu-\tau}$ charges of $\ell_4$ and $\ell_5$ are opposite in sign, the scale-dependent factor from  a renormalization scheme is cancelled, and the contributions of the vector-like leptons are basically similar to those of $\mu$ and $\tau$ leptons. In order to present the influence of $\tau'$ and $\tau''$, we show the $\epsilon$ as a function of $E_\gamma$ in Fig.~\ref{fig:km}, where the relation of $E_\gamma$ and $q^2$ is given by $E_\gamma=(s-q^2)/(2\sqrt{s})$; $\sqrt{s}$ is the center-of-mass energy of $e^+ e^-$, and $\sqrt{s}=10.58$ GeV is used. In the left panel,  the solid, dashed, dotted, and dot-dashed lines denote the results of $m_{L2}=(0.7,\,0.9,\,1.1,\, 1.5)$ TeV, respectively, and $m_{L1}=0.7$ TeV is fixed. The horizontal lines denote the same situations for the $\epsilon_{\nu e}$ results.  In the right panel, we fix $m_{L2}=0.7$ TeV and show the results with $m_{L1}=(0.7,\, 0.9,\, 1.1,\, 1.5)$ TeV for $\epsilon$ and $\epsilon_{\nu e}$.
We note that the results with $m_{L1}=m_{L2}=0.7$ TeV (solid) are the same as those without vector-like leptons.  
%
From the  plots, it can be seen that $\epsilon$ in the small $E_\gamma$ region (i.e., larger $m_{Z'}$) is sensitive to the ratio $m_{L2}/m_{L1}$. 
However, for the $e^+ e^- \to \gamma + \slashed{E}$ process, the SM backgrounds dominate in the small $E_{\gamma}$ region. To clearly understand how the $q^2=m^2_{Z'}$, $(m_{L_1}$, and $m_{L_2})$ affect the discovery significance, with the selected values of $\sqrt{q^2}=m_{Z'}$ and $(m_{L_1}, m_{L_2})$, we show the numerical values for the signal ($N_S$) and background ($N_B$) numbers and the corresponding significance, defined by $N_{S}/\sqrt{N_B + N_S}$, in Table~\ref{tab:event}, where $g_{Z'}=10^{-3}$ is fixed and the integrated luminosity of 50 ab$^{-1}$ is used in the numerical calculations; here we applied formulas for the signal and SM background cross section in Ref.~\cite{Araki:2017wyg}. 
It can be found that the significance can be over $3\sigma$ as $m_{Z'} \lesssim 1.0$ GeV and is increased(decreased) for $m_{L_1} < (>) \ m_{L_2}$. 

 \begin{figure}[tb]
\includegraphics[width=75mm]{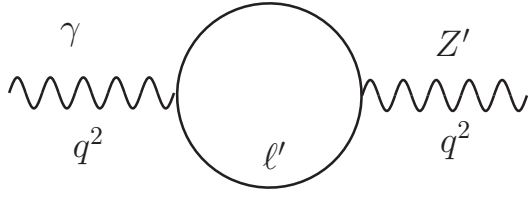} 
\caption{ Sketched Feynman diagram for the kinetic mixing of $\gamma$ and $Z'$, where the leptons inside the loop include $\ell'=\mu, \tau, \tau'$, and $\tau''$.}
\label{fig:km}
\end{figure}

 \begin{figure}[tb]
\includegraphics[width=75mm]{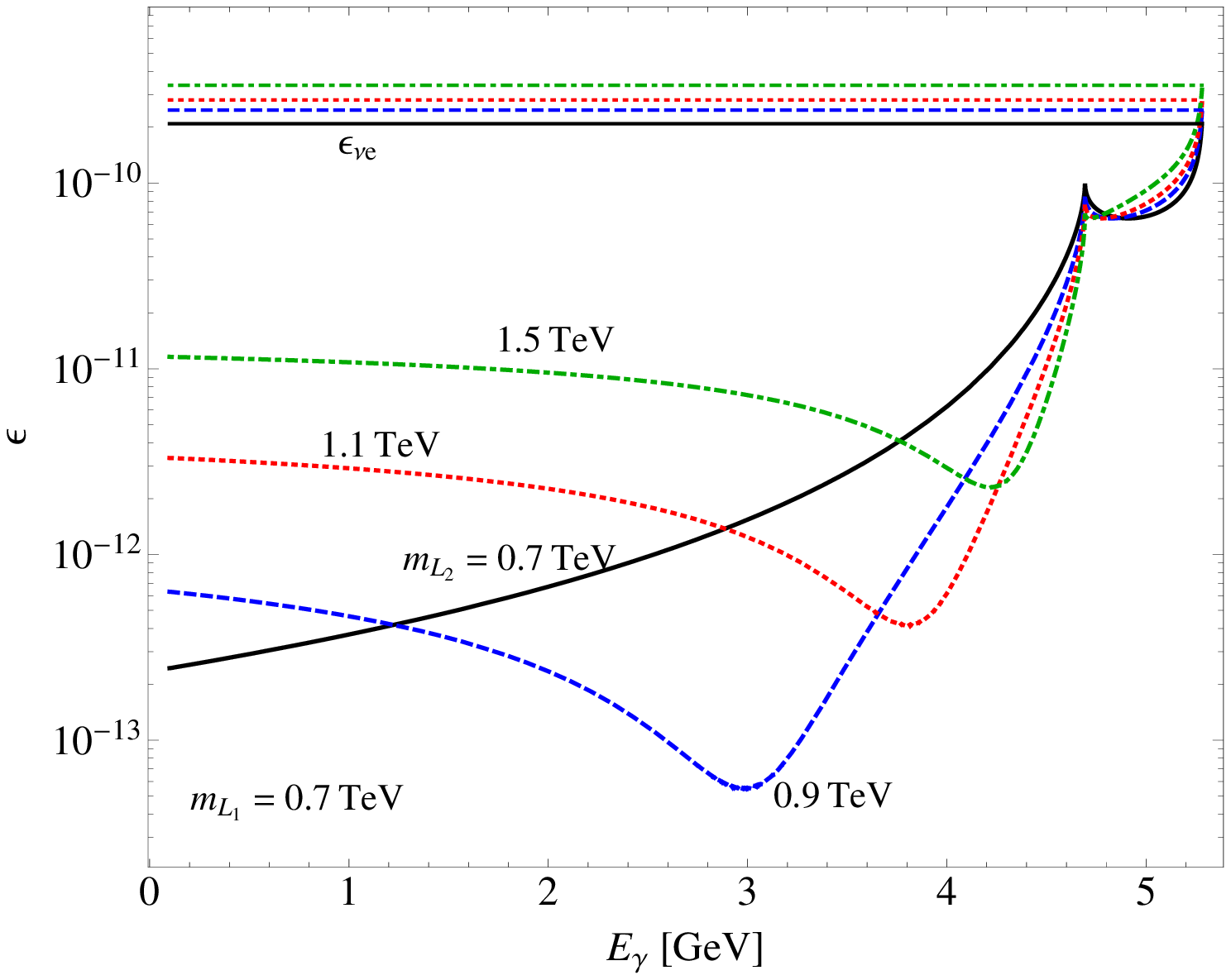} 
\includegraphics[width=75mm]{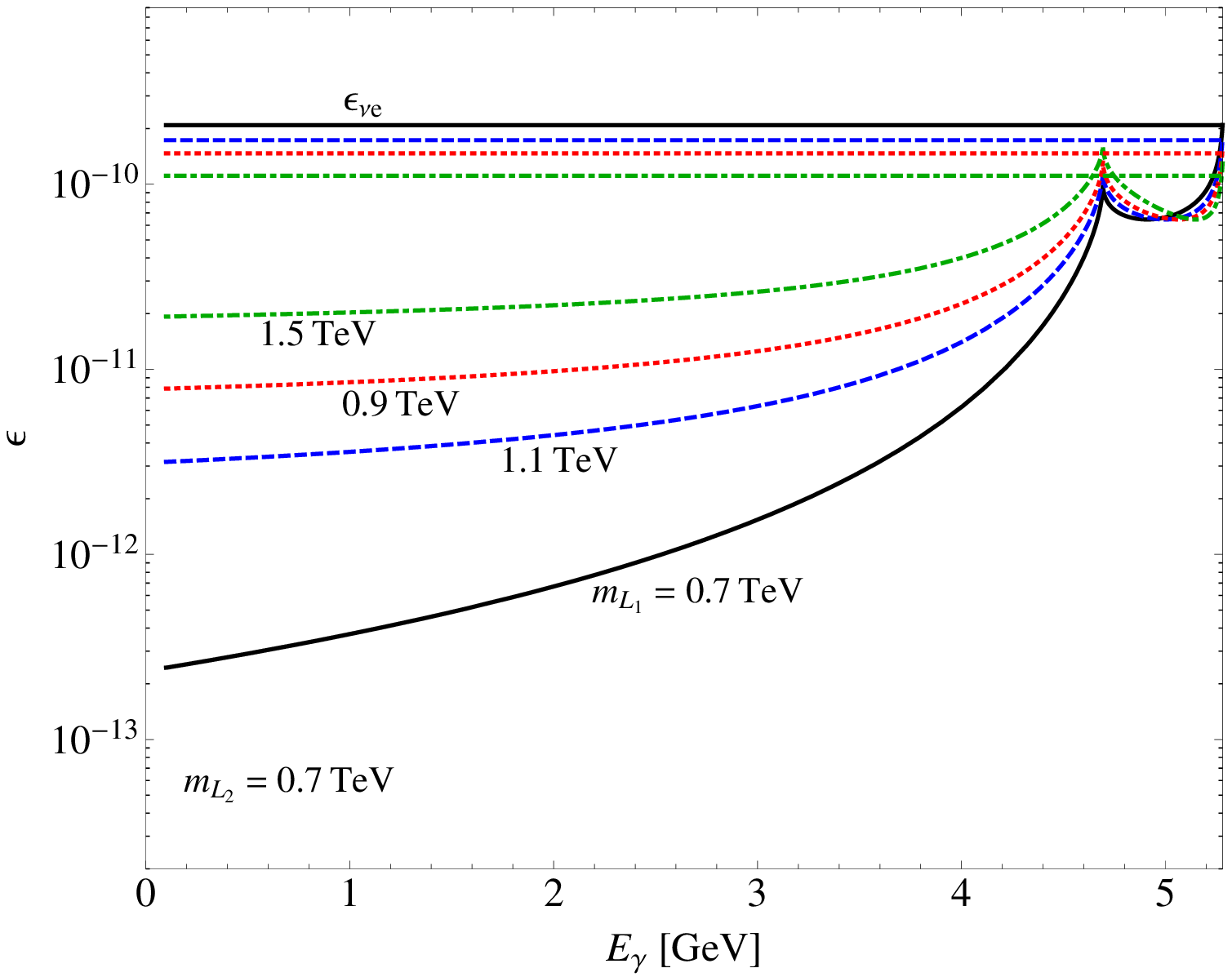} 
\caption{ Left panel: $\epsilon$  for $m_{L2} > m_{L1}$  in Eq.~(\ref{eq:Pi}) as a function of $E_\gamma$, where $E_\gamma=(s-q^2)/(2\sqrt{s})$, and  $\sqrt{s}=10.58$ GeV and  $m_{L1}=0.7$ TeV are used; the solid, dashed, dotted, and dot-dashed lines denote the results of $m_{L2}=(0.7, 0.9, 1.1, 1.5)$ TeV, respectively.  The horizontal  lines are the same situations but for the case of $\epsilon_{\nu e}=\Pi(0)$. The contributions of $m_{L1}=m_{L2}=0.7$ TeV are the same as for those without vector-like leptons. Right panel: the legend is the same as that in the left panel, but for $m_{L1}> m_{L2}$. }
\label{fig:km}
\end{figure}

\begin{table}[htp]
\caption{ Number of signal events $N_S$ and corresponding significance, defined by $N_S/\sqrt{N_S + N_B}$, for several taken values of $(m_{L_1}, m_{L_2})$,  where $g_{Z'}=10^{-3}$ is fixed;  the integrated luminosity of 50 ab$^{-1}$ is used, and the cases for $\sqrt{q^2} = m_{Z'}=(1.0, 1.5)$  GeV are presented.  The number of background events is  $N_B = 8 (30)$ for $m_{Z'}=1.0 (1.5)$ GeV.  }
\begin{tabular}{cccccc} \hline \hline
      ~~$(m_{L_1}, m_{L_2})$  [TeV] ~~   &  ~~$(0.7, 0.7)$~~  & ~~$(0.7, 1.1)$~~ & ~~ $(0.7, 1.5)$ ~~& ~~ $(1.1,0.7)$  ~~ & ~~ $(1.5,0.7)$  ~~\\ \hline 
 $N_S$ &   14 (11)   &   19 (14)      &  23 (17)      &  11 (9)  & 9 (8) \\ \hline 
 Significance &  3.0 (1.7)   &   3.6 (2.1)   &  4.1 (2.5)      &  2.5 (1.4)  & 2.2 (1.3)  \\\hline\hline
\end{tabular}
\label{tab:event}
\end{table}%


%


\subsection{Collider signatures} 

  Two heavy vector-like leptons $\tau'$ and $\tau''$ are introduced to generate the LFV scalar decays in this paper; therefore, it is of interest to study the  production of the heavy leptons  at the LHC. Since event simulation is beyond the scope of this paper, in the following we briefly discuss the potential channels and their production cross-sections. 

 In addition to the interactions of the $Z/\gamma$ gauge bosons and $\tau'(\tau'')$,  from Eqs.~(\ref{eq:ZI}), (\ref{eq:L_h}),  and (\ref{eq:L_phiS}), we also have the flavor-changing couplings,  expressed as:
 \begin{align}
&  \mu-\tau'-Z:~ \frac{g y y_\mu}{2\sqrt{2} c_W m_{L1}}\,, \quad  \mu-\tau''-Z:~ \frac{g y y'_\tau}{2\sqrt{2} c_W m_{L2}}\,, \nonumber \\
& \mu-\tau'-  h : ~\frac{y_\mu}{\sqrt{2}} \,, \quad  \quad  \quad  \quad ~~ \tau-\tau'-\phi_S :~ \frac{y_\tau}{\sqrt{2}} \,, \nonumber \\
&  \tau-\tau'' - h :~\frac{y'_\tau}{\sqrt{2}}\,, \quad  \quad \quad  \quad  ~~\mu-\tau'' -\phi_S:~ \frac{y'_\mu}{\sqrt{2}}\,.\label{eq:HLdecay}
 \end{align}
Therefore, the heavy leptons can be produced through the single and pair production channels. For singlet $\tau'(\tau'')$ production, the main processes in $pp$ collisions are from $q \bar q \to Z^* \to \tau' \mu (\tau'' \tau)$ and $gg\to h^* \to \tau' \mu (\tau'' \tau)$. The $\tau'(\tau'')$ pair production processes are via the $Z/\gamma$ gauge boson exchange. 
Using CalcHEP 3.6~\cite{Belyaev:2012qa} with CTEQ6 parton distribution functions (PDFs)~\cite{Nadolsky:2008zw}, the single $\tau'/\tau''$ and $\tau'/\tau''$ pair production cross-sections with $m_{\tau'(\tau'')}=(0.7, 0.9, 1.1)$ TeV and $y_\mu\sim y'_\tau\sim 1$ at $\sqrt{s}=13$ TeV are shown in Table~\ref{tab:CrX}, where we have used a K-factor of 1.5 for $gg\to h^* \to \tau' \mu (\tau'' \tau)$. We note that the values of the single and pair production cross-sections accidentally are the same due to  the use of $y_\mu\sim y'_\tau\sim 1$.
\begin{table}[htp]
\caption{ Single $\tau'/\tau''$ and $\tau'/\tau''$ pair production cross-sections at $\sqrt{s}=13$ TeV. }
\begin{tabular}{c|ccc} \hline \hline
   $m_{\tau'/\tau''}$ [TeV]    &  ~~~$0.7$~~~  & ~~~$0.9$~~~ & ~~~ $1.1$ ~~~ \\ \hline 
 $\sigma(\tau'/\tau'')$ [fb]&  0.43   &   0.11    &    0.036    \\ \hline 
 $\sigma(\tau'\tau'/\tau''\tau'')$ [fb] &  0.43   &  0.11   &  0.035      \\\hline\hline
\end{tabular}
\label{tab:CrX}
\end{table}%



 From Eq.~(\ref{eq:HLdecay}), it can be clearly seen that  the $\tau' \to \mu Z$ and $\tau''\to \tau Z$ decays have the suppression factors $v/m_{L1,L2}$; therefore, the heavy lepton decaying to the Higgs and $\phi_S$  are the dominant channels. If we assume $y_\mu\sim y_\tau$ and  $y'_\mu\sim y'_\tau$ and neglect the $m_{h,\phi_S}$ effects due to $m_{h,\phi_S} \ll m_{\tau',\tau''}$, the BRs can be simplified to be $BR(\tau'\to \mu h) \sim BR(\tau' \to \tau \phi_S) \sim BR(\tau''\to \tau h) \sim BR(\tau''\to \mu \phi_S) \sim 1/2$. Since $\phi_S$ has not yet been observed, the better discovery channels are through the Higgs production; accordingly, the collider signatures can be expressed as: 
 \begin{align}
 & pp \to \bar \tau' \mu ( \bar\tau'' \tau) \to \bar \mu \mu h ( \bar \tau \tau h) \,, \nonumber \\
 &   pp \to \bar \tau' \tau'  (\bar \tau'' \tau'') \to \bar\mu \mu h h ( \bar \tau \tau hh)\,.
 \end{align}
 With the luminosity of 300 fb$^{-1}$ and $m_{\tau'(\tau'')}=0.7$ TeV, the event numbers for the single and pair production are estimated as 65 and 32, respectively. Since the discovery significance  depends on the background events and kinematic analysis, we leave the detailed study for  future work.


\section{Summary}

We studied the $U(1)_{L_\mu -L_\tau}$ extension of the SM by including a pair of singlet vector-like leptons, where both heavy leptons carry different $L_\mu-L_\tau$ charges.  We employ a complex singlet scalar field to dictate the spontaneous $U(1)_{L_\mu - L_\tau}$ symmetry breaking. With $g_{Z'}\sim O(10^{-3})$,  the VEV of the singlet scalar field must be  at the electroweak scale in order to obtain $m_{Z'}$ in the MeV to GeV region. 
 It is found that the scalar boson contributions to the muon $g-2$ and the $h\to  \mu \tau$ decay  are strongly correlated in this model when the condition $a_L \simeq a_R$ is assumed. As a result, when $BR(h\to  \mu \tau)\sim 10^{-3}$ is taken,  the scalar-mediated muon $g-2$  can reach $10\times 10^{-10}$. Moreover, even with $g_{Z'}\sim 10^{-4}$, the muon $g-2$ combining $Z'$ and $\phi_S$ contributions can fit the current data with $2\sigma$ errors. The kinetic mixing in the $e^+ e^- \to \gamma + \slashed{E}$ process not only depends on the $q^2=m^2_{Z'}$, but also is sensitive to the ratio of $m_{L2}/m_{L1}$; as a result,  the significance of discovering the signal of $e^+ e^- \to \gamma \slashed{E}$ increases (decreases) for $m_{L2} > (<) \ m_{L1}$. 
It is found that $BR(\tau \to  \mu Z'Z')$ by the $\phi_S$ mediation can be of the ${\cal O}(10^{-8})$. When the BRs for $Z'\to (\bar \nu \nu, \mu^+ \mu^-)$ are included, $BR(\tau \to 3 \mu + \slashed{E}, 5\mu)$ of $10^{-9}$ can fall within the sensitivity of Belle II experiment in the search for the rare tau decays.  In addition, we briefly discuss the collider signatures for discovering the heavy leptons $\tau'$ and $\tau''$. The promising channels are through the Higgs production and given as $pp\to \bar \mu \mu (\bar \tau \tau)h$ and $pp\to \bar\mu \mu (\bar \tau \tau) hh$. \\

\noindent{\bf Acknowledgments}

This work was partially supported by the Ministry of Science and Technology of Taiwan,  under grant MOST-103-2112-M-006-004-MY3 (CHC).

\section*{\large Appendix} 

The Higgs Yukawa couplings in Eq.~(\ref{eq:LhS}) is written as:
\begin{align}
 -{\cal L}_{h} &=  \left(\begin{array}{cc}
  \pmb{\bar \ell}_L \,, & \pmb{\bar \Psi}_{\tau' L} \end{array} \right)  V_L \left( \begin{array}{c|c}
  ~~~\pmb{ m_{\ell} }_{3\times 3}~~~ &  \pmb{ \delta m_1 } \\ \hline
0 &0
 \end{array} \right) V^\dagger_R
  \left(\begin{array}{c}
   \pmb{\ell_R} \\
    \pmb{\Psi}_{\tau' R}  \end{array}\right) \frac{h}{v}  + H.c.\,, \nonumber \\
 & \approx   \left(\begin{array}{ccc}
  \pmb{\bar \ell}_L \,, & \pmb{\bar \Psi}_{\tau' L}  \end{array} \right)   \left( \begin{array}{c|c}
 \pmb{ m}_{\ell}  - \pmb{\delta m}_{1} \pmb{\epsilon}^\dagger_R &  \pmb{ \delta m_1 } \\ \hline
   0 & \pmb{\epsilon}^\dagger_L \pmb{\delta m}_{1}
 \end{array} \right)
  \left(\begin{array}{c}
   \pmb{\ell_R} \\
   \pmb{\Psi}_{\tau' R}  \end{array}\right) \frac{h}{v}  + H.c.\,,
      \end{align}
  where we have applied the flavor mixing matrices of Eq.~(\ref{eq:VI_II}) in the second line, and  $\pmb{\delta m}_1 \pmb{\epsilon}^\dagger_R$ and  $\pmb{\epsilon}^\dagger_L  \pmb{\delta m}_1$ are given by:
  \begin{equation}
  \pmb{\delta m}_1 \pmb{\epsilon}^\dagger_R = \frac{v v_S}{2}\left(\begin{array}{ccc}
  0 &0 & 0 \\
  0 & y_\mu y'_\mu \left( -\frac{c_\alpha s_\alpha}{m_{L1}}  + \frac{c_\alpha s_\alpha}{m_{L2}}\right)  &  y_\mu y_\tau\left( \frac{c^2_\alpha}{m_{L1}}  + \frac{s^2_\alpha}{m_{L2}}\right) \\
  0 & y'_\mu y'_\tau \left( \frac{s^2_\alpha}{m_{L1}}  + \frac{c^2_\alpha}{m_{L2}}\right) & y_\tau y'_\tau \left( -\frac{c_\alpha s_\alpha}{m_{L1}}  + \frac{c_\alpha s_\alpha}{m_{L2}}\right)
   \end{array} \right)  \,,
     \end{equation}
 \begin{equation}
  \pmb{\epsilon}^\dagger_L  \pmb{\delta m}_1 = \frac{v^2}{2}\left(\begin{array}{cc}
  \frac{y^2_\mu c^2_\alpha + y'^2_\tau s^2_\alpha}{m_{L1}} &  \frac{y^2_\mu -y'^2_\tau  }{m_{L1}}  c_\alpha s_\alpha \\
 \frac{y^2_\mu -y'^2_\tau  }{m_{L2}}  c_\alpha s_\alpha &   \frac{y^2_\mu s^2_\alpha + y'^2_\tau c^2_\alpha}{m_{L2}}  \end{array} \right)  \,.
 \end{equation}
Hence, the Higgs couplings to the charged leptons can be decomposed as:
 \begin{align}
 -{\cal L}_h & = \frac{m_\ell}{v} \bar \ell_L \ell_R h  + \frac{v_S  c_\alpha s_\alpha}{2} \frac{m_{L2} - m_{L1}}{m_{L1} m_{L2}} \left( y_\mu y'_\mu \bar \mu_L \mu_R  +  y_\tau y'_\tau \bar \tau_L \tau_R\right) h \nonumber \\
 & - \frac{v_S y_\mu y_\tau}{2}  \left( \frac{c^2_\alpha }{m_{L1}} + \frac{s^2_\alpha}{m_{L2}} \right) \bar \mu_L \tau_R h -   \frac{v_S y'_\mu y'_\tau}{2}  \left( \frac{s^2_\alpha }{m_{L1}} + \frac{c^2_\alpha}{m_{L2}} \right) \bar \tau_L \mu_R h \nonumber \\
 & + \frac{y_\mu}{\sqrt{2}} \bar \mu_L \left(c_\alpha \tau'_R  + s_\alpha \tau''_R \right) h + \frac{y'_\tau}{\sqrt{2}}  \bar\tau_L \left( -s_\alpha \tau'_R + c_\alpha \tau''_R \right) h + \pmb{\bar \Psi}_{\tau' L}  \pmb{\epsilon}^\dagger_L  \pmb{\delta m}_1  \pmb{\Psi}_{\tau' R} \frac{h}{v}+ H.c. \label{eq:L_h}
 \end{align}

 Similarly, the $\phi_S$ Yukawa couplings in Eq.~(\ref{eq:LhS})  is expressed as:
  \begin{align}
    -{\cal L}_{\phi_S } &=   \left(\begin{array}{cc}
 \pmb{\bar \ell}_L \,, & \pmb{\bar \Psi}_{\tau' L} \end{array} \right)  V_L  \left( \begin{array}{c|c}
  ~~~0~~~ & 0  \\ \hline
\pmb{\delta m^T_2} &0
 \end{array} \right)  V^\dagger_R
  \left(\begin{array}{c}
   \pmb{\ell_R} \\
    \pmb{\Psi}_{\tau' R}  \end{array}\right) \frac{\phi_S}{v_S}\,,  \nonumber \\
    & \approx  \left(\begin{array}{cc}
   \pmb{\bar \ell}_L \,, & \pmb{\bar \Psi}_{\tau' L} \end{array} \right)  V_L  \left( \begin{array}{c|c}
  - \pmb{\epsilon}_{L} \pmb{\delta m}^T_2 & 0  \\ \hline
\pmb{\delta m}^T_2 &  \pmb{\delta m}^T_2 \pmb{\epsilon}_R 
 \end{array} \right)  V^\dagger_R
  \left(\begin{array}{c}
   \pmb{\ell_R} \\
    \pmb{\Psi}_{\tau' R}  \end{array}\right) \frac{\phi_S}{v_S}\,,
     \end{align}
where $\pmb{\epsilon}_L  \pmb{\delta m}^T_2 = \pmb{\delta m}_1 \pmb{\epsilon}^\dagger_R$, and $\pmb{m}^T_2 \pmb{\epsilon}_R$ is written as:
  \begin{equation}
  \pmb{m}^T_2 \pmb{\epsilon}_R  = \frac{v^2_S}{2}\left(\begin{array}{cc}
  \frac{y'^2_\mu s^2_\alpha + y^2_\tau c^2_\alpha}{m_{L1}} &  \frac{-y'^2_\mu +y^2_\tau  }{m_{L2}}  c_\alpha s_\alpha \\
 \frac{-y'^2_\mu +y^2_\tau  }{m_{L1}}  c_\alpha s_\alpha &   \frac{y'^2_\mu c^2_\alpha + y^2_\tau s^2_\alpha}{m_{L2}}  \end{array} \right)   \,. 
 \end{equation}
Then, the $\phi_S$ Yukawa couplings to the charged leptons can be written as:
\begin{align}
 -{\cal L}_{\phi_S} & =   \frac{v  c_\alpha s_\alpha}{2} \frac{m_{L2} - m_{L1}}{m_{L1} m_{L2}} \left( y_\mu y'_\mu \bar \mu_L \mu_R  +  y_\tau y'_\tau \bar \tau_L \tau_R\right) \phi_S - \frac{v y_\mu y_\tau}{2}  \left( \frac{c^2_\alpha }{m_{L1}} + \frac{s^2_\alpha}{m_{L2}} \right) \bar \mu_L \tau_R \phi_S \nonumber \\
 &  -   \frac{v y'_\mu y'_\tau}{2}  \left( \frac{s^2_\alpha }{m_{L1}} + \frac{c^2_\alpha}{m_{L2}} \right) \bar \tau_L \mu_R \phi_S + \frac{y'_\mu}{\sqrt{2}}  \left( - s_\alpha \bar \tau'_L + c_\alpha \bar \tau''_L \right)  \mu_R \phi_S \nonumber \\
 &  + \frac{y_\tau}{\sqrt{2}} \left(c_\alpha \bar \tau'_L  + s_\alpha \bar \tau''_L \right) \tau_R \phi_S + \pmb{\bar \Psi}_{\tau' L}  \pmb{\delta m}^T_2 \pmb{\epsilon}_R   \pmb{\Psi}_{\tau' R} \frac{\phi_S}{v_S}  + H.c. \label{eq:L_phiS}
 \end{align}



\begin{thebibliography}{99}

\bibitem{He:1990pn} 
  X.~G.~He, G.~C.~Joshi, H.~Lew and R.~R.~Volkas,
  Phys.\ Rev.\ D {\bf 43}, 22 (1991).
  
\bibitem{Foot:1994vd} 
  R.~Foot, X.~G.~He, H.~Lew and R.~R.~Volkas,
  Phys.\ Rev.\ D {\bf 50}, 4571 (1994)
  [hep-ph/9401250].
  
\bibitem{Gninenko:2001hx} 
  S.~N.~Gninenko and N.~V.~Krasnikov,
  Phys.\ Lett.\ B {\bf 513}, 119 (2001)
  [hep-ph/0102222].
  
\bibitem{Gninenko:2014pea} 
  S.~N.~Gninenko, N.~V.~Krasnikov and V.~A.~Matveev,
  Phys.\ Rev.\ D {\bf 91}, 095015 (2015)
  [arXiv:1412.1400 [hep-ph]].
  
\bibitem{Altmannshofer:2016brv} 
  W.~Altmannshofer, C.~Y.~Chen, P.~S.~Bhupal Dev and A.~Soni,
  Phys.\ Lett.\ B {\bf 762}, 389 (2016)
  [arXiv:1607.06832 [hep-ph]].

\bibitem{Aartsen:2014gkd} 
  M.~G.~Aartsen {\it et al.} [IceCube Collaboration],
  Phys.\ Rev.\ Lett.\  {\bf 113}, 101101 (2014)
  [arXiv:1405.5303 [astro-ph.HE]].
  
\bibitem{Araki:2014ona} 
  T.~Araki, F.~Kaneko, Y.~Konishi, T.~Ota, J.~Sato and T.~Shimomura,
  Phys.\ Rev.\ D {\bf 91}, no. 3, 037301 (2015)
  [arXiv:1409.4180 [hep-ph]].

\bibitem{Kamada:2015era} 
  A.~Kamada and H.~B.~Yu,
  Phys.\ Rev.\ D {\bf 92}, no. 11, 113004 (2015)
  [arXiv:1504.00711 [hep-ph]].

\bibitem{DiFranzo:2015qea} 
  A.~DiFranzo and D.~Hooper,
  Phys.\ Rev.\ D {\bf 92}, no. 9, 095007 (2015)
  [arXiv:1507.03015 [hep-ph]].

\bibitem{Araki:2015mya} 
  T.~Araki, F.~Kaneko, T.~Ota, J.~Sato and T.~Shimomura,
  Phys.\ Rev.\ D {\bf 93}, no. 1, 013014 (2016)
  [arXiv:1508.07471 [hep-ph]].


\bibitem{Altmannshofer:2014cfa} 
  W.~Altmannshofer, S.~Gori, M.~Pospelov and I.~Yavin,
  Phys.\ Rev.\ D {\bf 89}, 095033 (2014)
  [arXiv:1403.1269 [hep-ph]].

\bibitem{Crivellin:2015mga} 
  A.~Crivellin, G.~D'Ambrosio and J.~Heeck,
  Phys.\ Rev.\ Lett.\  {\bf 114}, 151801 (2015)
  [arXiv:1501.00993 [hep-ph]].
  


\bibitem{Altmannshofer:2016jzy} 
  W.~Altmannshofer, S.~Gori, S.~Profumo and F.~S.~Queiroz,
  JHEP {\bf 1612}, 106 (2016)
  [arXiv:1609.04026 [hep-ph]].

\bibitem{Heeck:2014qea} 
  J.~Heeck, M.~Holthausen, W.~Rodejohann and Y.~Shimizu,
  Nucl.\ Phys.\ B {\bf 896}, 281 (2015)
  [arXiv:1412.3671 [hep-ph]].
  
\bibitem{Heeck:2016xkh} 
  J.~Heeck,
  Phys.\ Lett.\ B {\bf 758}, 101 (2016)
  [arXiv:1602.03810 [hep-ph]].

\bibitem{Altmannshofer:2016oaq} 
  W.~Altmannshofer, M.~Carena and A.~Crivellin,
  Phys.\ Rev.\ D {\bf 94}, no. 9, 095026 (2016)
  [arXiv:1604.08221 [hep-ph]].
  
  
\bibitem{Baek:2015mna} 
  S.~Baek, H.~Okada and K.~Yagyu,
  JHEP {\bf 1504}, 049 (2015)
  [arXiv:1501.01530 [hep-ph]].
  
\bibitem{Baek:2015fea} 
  S.~Baek,
  Phys.\ Lett.\ B {\bf 756}, 1 (2016)
  [arXiv:1510.02168 [hep-ph]].
  
\bibitem{Patra:2016shz} 
  S.~Patra, S.~Rao, N.~Sahoo and N.~Sahu,
  Nucl.\ Phys.\ B {\bf 917}, 317 (2017)
  [arXiv:1607.04046 [hep-ph]].
  
\bibitem{Biswas:2016yan} 
  A.~Biswas, S.~Choubey and S.~Khan,
  JHEP {\bf 1609}, 147 (2016)
  [arXiv:1608.04194 [hep-ph]].
  
\bibitem{Lee:2017ekw} 
  S.~Lee, T.~Nomura and H.~Okada,
  arXiv:1702.03733 [hep-ph].
  
  

  
\bibitem{Geiregat:1990gz} 
  D.~Geiregat {\it et al.} [CHARM-II Collaboration],
  Phys.\ Lett.\ B {\bf 245}, 271 (1990).

\bibitem{Mishra:1991bv} 
  S.~R.~Mishra {\it et al.} [CCFR Collaboration],
  Phys.\ Rev.\ Lett.\  {\bf 66}, 3117 (1991).

\bibitem{Altmannshofer:2014pba} 
  W.~Altmannshofer, S.~Gori, M.~Pospelov and I.~Yavin,
  Phys.\ Rev.\ Lett.\  {\bf 113}, 091801 (2014)
  [arXiv:1406.2332 [hep-ph]].

\bibitem{TheBABAR:2016rlg} 
  J.~P.~Lees {\it et al.} [BaBar Collaboration],
  Phys.\ Rev.\ D {\bf 94}, no. 1, 011102 (2016)
  [arXiv:1606.03501 [hep-ex]].


\bibitem{Harnik:2012ni} 
  R.~Harnik, J.~Kopp and P.~A.~N.~Machado,
  JCAP {\bf 1207}, 026 (2012)
  [arXiv:1202.6073 [hep-ph]].

\bibitem{Bellini:2011rx} 
  G.~Bellini {\it et al.},
  Phys.\ Rev.\ Lett.\  {\bf 107}, 141302 (2011)
  [arXiv:1104.1816 [hep-ex]].
  
\bibitem{Kaneta:2016uyt} 
  Y.~Kaneta and T.~Shimomura,
  arXiv:1701.00156 [hep-ph].

\bibitem{Araki:2017wyg} 
  T.~Araki, S.~Hoshino, T.~Ota, J.~Sato and T.~Shimomura,
  Phys.\ Rev.\ D {\bf 95}, no. 5, 055006 (2017)
  [arXiv:1702.01497 [hep-ph]].




\bibitem{Aushev:2010bq} 
  T.~Aushev {\it et al.},
  arXiv:1002.5012 [hep-ex].




\bibitem{Bennett:2006fi} 
  G.~W.~Bennett {\it et al.} [Muon g-2 Collaboration],
  Phys.\ Rev.\ D {\bf 73}, 072003 (2006)
  [hep-ex/0602035].

\bibitem{Grange:2015fou} 
  J.~Grange {\it et al.} [Muon g-2 Collaboration],
  arXiv:1501.06858 [physics.ins-det].
  
\bibitem{Otani:2015jra} 
  M.~Otani [E34 Collaboration],
  JPS Conf.\ Proc.\  {\bf 8}, 025008 (2015).


\bibitem{Aad:2016blu} 
  G.~Aad {\it et al.} [ATLAS Collaboration],
  Eur.\ Phys.\ J.\ C {\bf 77}, no. 2, 70 (2017)
  [arXiv:1604.07730 [hep-ex]].


\bibitem{Khachatryan:2015kon}
  V.~Khachatryan {\it et al.} [CMS Collaboration],
  Phys.\ Lett.\ B {\bf 749} (2015) 337
  [arXiv:1502.07400 [hep-ex]].


\bibitem{CMS:2017onh} 
  CMS Collaboration [CMS Collaboration],
  CMS-PAS-HIG-17-001.



\bibitem{PDG} C. Patrignani et al. (Particle Data Group), Chin. Phys. C, {\bf 40}, 100001 (2016). 

\bibitem{Flores-Tlalpa:2015vga} 
  A.~Flores-Tlalpa, G.~Lopez Castro and P.~Roig,
  JHEP {\bf 1604}, 185 (2016)
  [arXiv:1508.01822 [hep-ph]].

      \bibitem{Belyaev:2012qa} 
  A.~Belyaev, N.~D.~Christensen and A.~Pukhov,
  Comput.\ Phys.\ Commun.\  {\bf 184}, 1729 (2013)
  [arXiv:1207.6082 [hep-ph]].

\bibitem{Nadolsky:2008zw} 
  P.~M.~Nadolsky, H.~L.~Lai, Q.~H.~Cao, J.~Huston, J.~Pumplin, D.~Stump, W.~K.~Tung and C.-P.~Yuan,
  Phys.\ Rev.\ D {\bf 78}, 013004 (2008)
  [arXiv:0802.0007 [hep-ph]].

\end{thebibliography}
\end{document}